\documentclass[12pt]{iopart} 
\usepackage{graphicx,latexsym,stmaryrd,iopams}

\newcommand{\nc}{\newcommand}
\nc{\be}{\begin{equation}}
\nc{\ee}{\end{equation}}
\nc{\beq}{\begin{eqnarray}}
\nc{\eeq}{\begin{eqnarray}}

\def\st[#1]{
| #1 \rangle
}

\begin{document}

\title[Spin-1 chain compounds]{Thermal and magnetic properties of integrable spin-1 and 
spin-$\frac32$ chains with applications to real compounds}

\author{M. T. Batchelor$^{\dagger}$, X.-W. Guan$^{\dagger}$, N. Oelkers$^{\dagger}$ 
and A. Foerster$^{\ddagger}$}
\address{$^{\dagger}$ Department of Theoretical Physics, Research School of Physical
Sciences \& Engineering, and Mathematical Sciences Institute,\\
Australian National University, Canberra ACT 0200,  Australia\\
$^{\ddagger}$ Instituto de Fisica
da UFRGS, Av.\ Bento Goncalves, 9500, \\Porto Alegre, 91501-970,
Brazil
}

\begin{abstract}
The ground state and thermodynamic properties of spin-$1$ and 
spin-$\frac32$ chains are investigated via exactly solved $su(3)$ and $su(4)$ 
models with physically motivated chemical potential terms. 
The analysis involves the Thermodynamic Bethe Ansatz and the 
High Temperature Expansion (HTE) methods.
For the spin-$1$ chain with large single-ion anisotropy, a gapped phase occurs
which is significantly different from the valence-bond-solid Haldane phase.
The theoretical curves for the magnetization, susceptibility and specific heat
are favourably compared with experimental data for a number of  
spin-1 chain compounds.
For the spin-$\frac32$ chain a degenerate gapped phase exists
starting at zero external magnetic field. 
A middle magnetization plateau can be triggered by the single-ion
anisotropy term.
Overall, our results lend further weight to the applicability of integrable models to the
physics of low-dimensional quantum spin systems.
They also highlight the utility of the exact HTE method. 
\end{abstract}



\maketitle

\tableofcontents

\newpage

\section{Introduction}
\label{sec:intro}

This paper is concerned with the study of two integrable quantum spin chains.
By integrable we mean that the chains are exactly solvable in the 
non-diffractive \cite{McG} Yang-Baxter 
sense \cite{Yang, Baxter}, with infinitely many conserved quantities and 
an underlying Bethe Ansatz solution.
The two particular models are variants of the spin-1 and spin-$\frac32$ chains based on the 
family of integrable su($n$) permutator models \cite{Uimin, Suth}.
The two-dimensional classical lattice model counterparts are the $A_{n-1}$ models \cite{VVB,Jimbo}, 
aka the Perk-Schultz model \cite{Perk} in a special case.

Over the past few decades considerable effort has been invested in the analysis of 
exactly solved models in statistical mechanics \cite{Bbook,J-M}.
Of particular relevance here is the calculation of the 
temperature-dependent properties  \cite{Tbook}.
This began with the Thermodynamic Bethe Ansatz (TBA)  approach \cite{Yang-Yang, TBA} and has most
recently evolved into the exact High Temperature Expansion (HTE) method \cite{HTE2, ZT}.
The key ingredients of the HTE method are the Quantum Transfer Matrix 
(QTM) \cite{QTMrefs, QTMproof} and the T-system \cite{KunibaFusion}, 
from which one derives nonlinear integral equations which can be 
solved in an exact perturbative fashion.
To date this approach has been applied to the Heisenberg model \cite{HTE2}, 
the $su(n)$ models \cite{ZT}, the higher spin Heisenberg model \cite{ZT2} and to integrable 
quantum spin ladders \cite{HTE1,YRFC}.
In particular, it has been demonstrated  that integrable models can be used to 
study real ladder compounds \cite{HTE1}. 
On the one hand, the TBA method was seen to be most convenient for
predicting critical fields \cite{TBA2,mix,ying} of relevance to
the quantum phase transitions in ladder compounds.
On the other hand, the HTE method gave the temperature-dependent 
free energy, from which physical properties such as the magnetic susceptibility and
the specific heat were derived.
Excellent agreement between the theoretical and
experimental results for the strong coupling ladder compounds  has been found  \cite{HTE1}.

Here we apply the TBA and  HTE methods to calculate the  thermal and magnetic properties of
spin-1 and spin-$\frac32$ chains.\footnote{A brief account of our results can be found in 
Ref.~\cite{BGO}.}
%

\subsection{Physical motivation}

Low-dimensional spin systems continue to reveal rich and novel quantum magnetic effects, 
such as fractional magnetization plateaux \cite{LADD1,FPL} and spin-Peierls transitions \cite{SPT}.
The effects of inter- and intra-chain interactions, single-ion anisotropy,
biquadratic exchange interactions and the magnetic field all provide
competing contributions in the low-temperature physics.
As an example consider the spin-$S$ Heisenberg chain \cite{Hald}:
if $2S$ is odd, the next-nearest-neighbour exchange interaction and
anisotropy may drive the chain into either a
non-magnetic singlet ground state (spin-Peierls transition) \cite{SPT}
or a magnetic gapped ground state \cite{AFF3}. 
Conversely, if $2S$ is even the Haldane gap may be closed in the
presence of additional biquadratic terms or in-plane anisotropy.  
In general, the Haldane phase exists only in $2S$ even Heisenberg chains
with antiferromagnetic coupling, no gap appears in ferromagnetic
Heisenberg chains.
Accordingly one expects a rich quantum phase
diagram to emerge for low-dimensional spin chains. 
More specifically, the intra- and inter-chain exchange interactions may result in either a
valence-bond-solid Haldane phase or a dimerized phase \cite{AFF1}, whereas
large single-ion  anisotropy may trigger a field-induced gapped phase \cite{AFF3,Tsvelik} 
significantly different from the standard Haldane phase.

The  spin-1 Heisenberg chain has been
extensively studied in the context of Haldane gapped materials 
\cite{Affleck1,NINO1,SP1C1,SP1C2, Hagi}.
Recently an anomaly in the susceptibility has been observed in the spin-1
compounds LiVGe$_2$O$_6$ \cite{Gavi}.
Such new materials have drawn attention from both theorists and 
experimentalists \cite{Lum, Von, Lou}.
The Heisenberg spin-1 chain with additional biquadratic interaction has also been studied with
regard to the Haldane phase \cite{Lou,Scholl}.
In particular, the susceptibility of LiVGe$_2$O$_6$ can be quantitatively explained with the
spin-1 bilinear-biquadratic chain \cite{Gavi,Lou}.
On the other hand, a field-theoretic approach to the Heisenberg spin-$1$ chain with 
single-ion anisotropy was suggested by Tsvelik \cite{Tsvelik}. 
His theoretical predictions are in good agreement with experimental results
for the compound Ni(C$_2$H$_8$N$_2$)NO$_2$ClO$_4$.  

\subsection{Application to real compounds}

The Haldane gapped phase is observed in some one-dimensional materials exhibiting
a weak single-ion anisotropy $D$, such that $D\leq J$, where $J$ is the nearest-neighbour
exchange interaction.  
However, a large anisotropy ($D> J$) can also produce a gapped phase \cite{AFF3},
which is significantly different from the Haldane phase found in  weakly anisotropic
compounds. 
In this latter case, the Haldane nondegenerate ground state is a valence-bond solid 
state \cite{AFF1} in which a single valence bond connects each neighbouring pair to form a singlet. 
An expected excitation from the valence-bond solid state arises from a configuration in 
which  a nonmagnetic state $S_i=0$ at site $i$ is substituted for a state with $S_i=1$. 
In this way  a total spin-1 excitation creates an energy referred to as the energy gap.  
However, bond alternation may lead to a dimerized ground state in which double valence 
bonds form two singlets between pairs of neighboring spins. 
By breaking one dimer with a nonmagnetic state at site $i$, a total spin-1 excitation is also
separated.

On the other hand, the field-induced gapped phase in the spin-1
chain is caused by trivalent orbital splitting, i.e., the singlet
and doublet orbitals are separated by a crystal-field splitting. 
The singlet can occupy all states such that the ground state lies in the
nondegenerate gapped phase.  
The lowest excitation arises as the doublet becomes involved in the ground state.  
This gapped phase is observed in some Nickel salts with large zero-field splitting, 
such as
NiSnCl$_6\cdot 6$H$_2$O \cite{PRB3488},
[Ni(C$_5$H$_5$NO)$_6$](ClO$_4$)$_2$ \cite{PRB3523} and 
Ni(NO$_3$)$_2\cdot 6$H$_2$O \cite{PRB4009}.  
Recently, other  spin-$1$ magnetic compounds:  
\begin{enumerate}
\item Ni(C$_2$H$_8$N$_2$)$_2$Ni(CN)$_4$ (abbreviated NENC)\cite{NENC,sus,ESR}, 
\item Ni(C$_{11}$H$_{10}$N$_2$O)$_2$Ni(CN)$_4$ (abbreviated as NDPK) \cite{NENC}, and
\item Ni(C$_{10}$H$_8$N$_2$)$_2$Ni(CN)$_4$$\cdot$H$_2$O (abbreviated NBYC)
\cite{NENC,NBYC},
\end{enumerate}
have also been identified as effective Heisenberg magnetic chains.  
This class of compounds exhibits a nondegenerate ground
state which can be separated from the lowest excitation. 
Theoretical studies of these compounds have relied on a
molecular field approximation via  the Van Vleck equation \cite{Carlin}. 
To first order Van Vleck approximation the exchange interaction
is neglected and an effective crystalline field is incorporated to fit
the experimental data.
Not surprisingly these approximations cause some discrepancies in fitting the data. 
Although there has been some theoretical interpretation, the nature of the quantum
phase transitions as well as the general microscopic Hamiltonian remains to be fully
clarified for these compounds. 

\subsection{Results and outline of this paper}

In this paper we consider integrable one-dimensional
spin-$1$ and spin-$\frac32$ models with large planar single-ion anisotropy. 
We find that for the integrable spin-$1$ chain the
non-magnetic singlet and the magnetic doublet states can be separated
from the lowest magnon excitation by an energy gap. 
 A large single-ion anisotropy together with in-plane anisotropy may trigger a
non-magnetic mid-plateau, leading to a ``ferrimagnetic-antiferromagnetic" phase transition. 
Significantly, a different type of gapped phase compared to the standard Haldane phase is found. 
We examine some real spin-$1$ compounds  via  TBA and HTE. 
Excellent agreement between our theoretical predictions and the
experimental results for  spin-$1$ compounds such as Nickel salts and 
magnetic chains (NENC, NBYC and NDPK) is found. 
Our results show that the strong single-ion anisotropy, which is induced by an orbital
splitting, can dominate the behaviour of these  compounds.

For the integrable spin-$\frac32$ chain the magnetic gapped ground state can be separated 
from the lowest magnon excitation, in contrast to the non-magnetic Haldane gapped phase in 
the spin-$1$ chain.
 A  magnetization plateau at $M=\frac12$ originates at zero magnetic field. 
Moreover, an appropriately chosen value of the anisotropy can open a
one third  magnetization plateau reminiscent of the mixed
spin-(1,$\frac12$) ladder model \cite{mix}.

The paper is organized as follows.
In \sref{sec:theory} we introduce the integrable su(3) spin-$1$ chain and briefly review 
its exact solution. 
Then in \sref{sec:tba} we discuss the ground state properties via the TBA.
In \sref{sec:hte} we derive the high temperature magnetic properties via the HTE method. 
We then discuss the application of the TBA and the HTE methods to real spin-$1$ materials
with single-ion and in-plane anisotropies in \sref{sec:exp}.
In particular, we examine the thermal and magnetic properties of the Nickel salt 
NiSnCl$_6\cdot 6$H$_2$O (\sref{sec:Nisalt}) and the magnetic chains
NENC (\sref{sec:NENC}), NBYC (\sref{sec:NBYC}) and NDPK (\sref{sec:NDPK}) by
comparing our TBA and HTE results with the 
experimental data for the specific heat, magnetic susceptibiliy and
magnetization, where available.
In \sref{sec:spin32} we apply the same approach to the su(4) case in the context of the
integrable spin-$\frac32$ chain.
Concluding remarks are given in  \sref{sec:conclusions}.

\section{The integrable spin-$1$ chain}
\label{sec:theory}

\subsection{The Heisenberg spin-$1$ chain}

Initially, the Haldane gap was identified in some quasi one-dimensional spin-$1$ materials
\cite{NINO1,SP1C1,SP1C2} described by the Heisenberg Hamiltonian
\begin{equation}
{\cal H}=J\sum_{j=1}^{L}\vec{S}_j\cdot
\vec{S}_{j+1}+D\sum_{j=1}^L(S_j^z)^2-\mu_B g H\sum_{j=1}^LS^z_j
\label{eq:HBchain}
\end{equation}
in a parallel external magnetic field $H$.
Here $\vec{S}_i$ denotes the spin-$1$ operator at site $i$ and $L$ is the number of sites. 
The constants $J$ and $D$ denote the exchange spin-spin coupling strength and 
single-ion anisotropy, respectively. 
The Bohr magneton  is denoted  by $\mu_B$ and $g$ is the Lande factor. 
For the compound NENP  \cite{SP1C2} the energy gap $\Delta$ between the
ground state and the first excited state is of 
magnitude $\Delta \approx 12 $ K  with the
coupling constants $J=48$ K and $D=0.2 J$. 
While for the compound NINO \cite{NINO1}, 
$\Delta\approx10-15 $ K with $J=52$ K and $D=0.3 J$.  
The above Hamiltonian has been  recognized as a good model for Haldane-like compounds 
and has thus found widespread interest in the theoretical  
community \cite{LADD1,Milla,HA}. 
Unfortunately it does not appear to be integrable.
However, if the spin-spin interaction term is changed, the model is integrable,
and thus amenable to exact calculations using the sophisticated and well developed machinery
of  integrable  models. 
We shall see below that under favourable conditions
the modified Hamiltonian exhibits similar physical behaviour.

\subsection{The integrable su(3) spin-1 chain}

We consider the Hamiltonian
\begin{eqnarray}	
{\cal H} = J\sum_{j=1}^L
\Bigg(\!\vec{S}_j\!\cdot\!\vec{S}_{j+1}\!+\!\!\left(\vec{S}_j\!\cdot\!\vec{S}_{j+1}\!\right)^{\!\!2}
\Bigg)
+{\cal H}_{{\textrm{\tiny chem.pot}}}
\label{eq:Ham1}
\end{eqnarray}
which only differs from the Heisenberg Hamiltonian \eref{eq:HBchain} in
the additional biquadratic interaction term.\footnote{There are in fact three 
integrable $su(2)$-invariant spin-1 Hamiltonians, each involving biquadratic 
interactions \cite{classify}. Only the model considered here, with two-body interactions based 
on the permutation operator, appears to allow the desired chemical potential 
terms.} 
We discuss the on-site chemical potential terms $ {\cal H}_{{\textrm{\tiny chem.pot}}}$ 
in detail further below.
Leaving aside $ {\cal H}_{{\textrm{\tiny chem.pot}}}$, the model \eref{eq:Ham1} is the 
integrable $su(3)$ spin-$1$ chain, which is well understood \cite{classify,Fujii}.
It belongs to a family of  solvable models with Lie algebra  $su(n)$ symmetry, 
which are generalized multi-state vertex models \cite{Jimbo,Martins}.
The spin-1 chain was originally considered by Uimin \cite{Uimin} in 1970 and later in the
context of multi-state permutation operators by Sutherland \cite{Suth}.
A key ingredient is the identity \cite{Uimin}
\be
P_{i,j} = \vec{S}_i\cdot\vec{S}_j +
\left(\vec{S}_i\cdot\vec{S}_j\right)^2 -{\bf I}_i\otimes {\bf I}_j, 
\label{eq:permutationidentity}
\ee 
where ${\bf I} $ is the identity operator.
We are thus considering the model
 \be
 {\cal H} = J \sum_{i=1}^L P_{i,i+1} +{\cal H}_{{\textrm{\tiny chem.pot}}}
 \label{eq:eq:Ham1}
 \ee
acting on $V^{\otimes L}$ where $V$ is a three-dimensional
vector space and $P_{i,i+1}$ is the permutation operator, acting 
trivially on all sites, except sites $i$ and $i+1$, where 
\be P_{i,i+1}
 \st[j_1,j_2,\ldots,\underbrace{j_i,j_{i+1}},\ldots,j_L] =
 \st[j_1,j_2,\ldots,\underbrace{j_{i+1},j_{i}},\ldots,j_L] .
\ee 
We take the spin-spin exchange interaction $J$ to be antiferromagnetic ($J>0$).

A crucial point in the following TBA and HTE analysis is the observation that the form 
of the permutation operator is not dependent on which basis spans $V$. 
Accordingly we always choose a local basis which diagonalizes the chemical 
potential terms. 
The chemical potentials characterize physical on-site interaction terms in the model Hamiltonian.
Fortunately there is freedom to adjust them without losing integrability.
  
To simplify the Bethe Ansatz, we apply periodic boundary conditions,
i.e., we identify $P_{L,L+1} = P_{L,1}$ and $V_{L+1} = V_1$.
Following the standard approach this model can be solved by the nested Bethe 
Ansatz \cite{Uimin,Suth}, which gives the energy eigenvalues 
\label{page:BetheAnsatz1}
\begin{equation}
{\cal E}=-J\sum_{j=1}^{M_1}\frac{1}{(v_j^{(1)})^2+\frac{1}{4}}
+ N_1\, \mu_1
+ N_2\, \mu_2
+ N_3\, \mu_3
\label{eq:Eeigenvalue}
\end{equation}
in terms of Bethe roots of flavour/colour $v_j^{(i)}$.
Only the Bethe roots of first flavour $v_j^{(1)}$appear in the energy expression. 
The roots are determined by solutions to the Bethe equations
\begin{eqnarray}
\left(\frac{v_j^{(1)}+\frac12\mathrm{i}}
{v_j^{(1)}-\frac12\mathrm{i}}\right)^L
&=\prod^{M_1}_{\stackrel{\scriptstyle l=1}{l\neq j}}
\frac{v_j^{(1)}-v_l^{(1)}+\mathrm{i}}{v_j^{(1)}-v_l^{(1)}-\mathrm{i}}
\prod_{l=1}^{M_{2}}\frac{v_j^{(1)}-v_l^{(2)}-\frac12\mathrm{i}}
{v_j^{(1)}-v_l^{(2)}+\frac12\mathrm{i}},
\nonumber \\
\prod_{i=1}^{M_{1}}\frac{v_k^{(2)}-v_i^{(1)}+\frac12\mathrm{i}}
{v_k^{(2)}-v_i^{(1)}-\frac12\mathrm{i}}
&=\prod^{M_2}_{\stackrel{\scriptstyle l=1}{l\neq k}}
\frac{v_k^{(2)}-v_l^{(2)}+\mathrm{i}}{v_k^{(2)}-v_l^{(2)}-\mathrm{i}},
\label{eq:BAEsu3}
\end{eqnarray}
where $ j=1, 2, \ldots, M_1$ and $k= 1, 2, \ldots, M_2$.

In eq.~\eref{eq:Eeigenvalue} $N_i$ denotes
the occupation number of sites in state $\st[i]$ for $\,i=1,2,3$, where the basis
$\st[1]$,$\st[2]$,$\st[3]$ diagonalizes the chemical potential term under
consideration with eigenvalues $\mu_1$, $\mu_2$, $\mu_3$. 
The Bethe quantum numbers $M_1$ and $M_2$, satisfying $0 \leq M_2 \leq M_1 \leq L$, 
are related to the number of sites in state $\st[i]$ by
\begin{eqnarray}
N_1 &=&\,L\ -M_1,\nonumber \\
N_2 &=&M_1-M_2,\label{eq:BANMrelation} \\
N_3 &=&M_2.\nonumber 
\end{eqnarray}
The solution \eref{eq:Eeigenvalue} in terms of the Bethe equations \eref{eq:BAEsu3}
is valid for any permutation of the basis order.
We do not specify the basis here, rather we are free to choose the {\sl order} of the basis and
we will exploit this in the TBA analysis of the ground state, as done in the TBA analysis
of the integrable ladder models \cite{TBA2,mix,ying,sun}. 

\subsubsection{Chemical potential terms}
 
We choose the chemical potential terms which most closely model the
physics of the compounds as used in the literature. 
The most general chemical potential terms exhibited in the real compounds are of the form
 \begin{eqnarray}
{\cal  H}_{{\textrm{\tiny chem.pot}}} &= -\mu_B g H_z \sum_{j=1}^{L}{S}^z_j + D \sum_{j=1}^{L}
({S}^z_j)^2 + E\sum_{j=1}^{L} \left(({S}^x_j)^2 -({S}^y_j)^2\right),
\label{eq:Chempotparallel}
\\
{\cal H}_{{\textrm{\tiny chem.pot}}} &= -\mu_B g H_x \sum_{j=1}^{L}{S}^x_j + D \sum_{j=1}^{L}
({S}^z_j)^2 + E \sum_{j=1}^{L}\big( ({S}^x_j)^2 -({S}^y_j)^2\big). 
\label{eq:Chempotperpendicular}
\end{eqnarray}
The first term in each equation describes the effect of an external magnetic field $H$, 
which can be applied either parallel or perpendicular to the quantization axis 
given by the sample geometry.   
The in-plane anisotropy and planar anisotropy effects may be detected in magnetic properties 
of a single crystal. 
Actually,  we can transform the transverse external
magnetic field $H_x$ in \eref{eq:Chempotperpendicular} to the $z$ direction via the 
following equivalent chemical potentials
\begin{eqnarray}
{\cal H}_{{\textrm{\tiny chem.pot}}} &= -\mu_B g H_z \sum_{j=1}^{L}{S}^z_j + D \sum_{j=1}^{L}
({S}^x_j)^2 + E \sum_{j=1}^{L}\big( ({S}^z_j)^2 -({S}^y_j)^2\big). 
\label{eq:Chempotperpendicular-equ}
\end{eqnarray}
Therefore, the magnetic field perpendicular to the quantized axis
provides a way to examine thermodynamic properties for the powdered
samples via the chemical potentials
\eref{eq:Chempotperpendicular-equ}.
We will assume that $D>0$ and interpret it as a planar/single-ion anisotropy modelling the 
effect of crystal-field splitting.  
The third contribution describes an in-plane anisotropy effect which breaks the $z^2 $ symmetry.
We consider only the case $E\geq 0$.

Although we do not need to consider the (complicated) eigenbasis of the chemical potential
terms \eref{eq:Chempotparallel} or \eref{eq:Chempotperpendicular} in
performing the algebraic Bethe ansatz and applying the HTE method, the eigenbasis for 
different chemical potentials is helpful in analyzing the ground state properties of the 
Hamiltonian \eref{eq:Ham1}.  
The eigenbasis could for instance be presented as linear combinations of
the spin projection onto the $z$-axis eigenstates $\st[S^z=\pm1,0]$
where the coefficients are dependent on $H$, $D$ and $E$. 
The eigenbasis for the Hamiltonian \eref{eq:Ham1} with different
chemical potentials will be given in the following  TBA
analysis. 
The eigenvalues of the chemical potentials are given by
\numparts
\begin{eqnarray}
 \label{eq:EV}
\mu_1&=& -\sqrt{E^2 +(\mu_B g H)^2},\label{eq:EV.1}\\
\mu_2&=& -D,\label{eq:EV.2}\\
\mu_3&=&\sqrt{E^2 +(\mu_B g H)^2},\label{eq:EV.3}\\
\mu_1&=& -\frac{1}{2}\left(D-E +\sqrt{(D+E)^2 +(2\mu_B g H)^2}\right),\label{eq:EV.4}\\
\mu_2&=& -E, \label{eq:EV.5}\\
\mu_3&=& -\frac{1}{2}\left(D-E- \sqrt{(D+E)^2 +(2\mu_B g H)^2}\right)
\label{eq:EV.6}
\end{eqnarray}
\endnumparts
for \eref{eq:Chempotparallel} and \eref{eq:Chempotperpendicular}, respectively.

Note that we can adjust the chemical potential terms to model the
physics of a specific compound without losing integrability, but we
have no freedom in changing the spin-spin exchange interaction fixed by the
permutation identity \eref{eq:permutationidentity}.  
In this way, the integrable $su(3)$ chain \eref{eq:Ham1} may describe some real 
compounds having either a spin-spin interaction close to the permutation interaction or 
exhibiting a strong single-ion anisotropy which effectively dominates the
intrachain exchange interaction in the ground state. 
Fortunately, compounds are known  where the relative strength $J/D$ or $J/E$
are very small. 
Thus the planar and in-plane anisotropies tend to dominate the physics.

A large $D$ anisotropy may trigger a gapped phase where the singlet
occupies the  whole ground state. 
In such a phase, the Bethe reference state becomes the true physical ground
state, with Bethe  numbers $M_1$ and $M_2$ in the Bethe equations
\eref{eq:BAEsu3} equal to zero.
The lowest excitation arises from a configuration in which the 
doublet is involved in the ground state. 
This excitation gives the energy gap, which can be
caculated either from the energy \eref{eq:Eeigenvalue} in terms of 
the Bethe equations \eref{eq:BAEsu3} or from the TBA equations.
We shall see that both the magnetic field and the in-plane anisotropy
 decrease the energy gap.

\subsection{TBA analysis: Ground state properties}
\label{sec:tba}

We consider the thermodynamic limit in which 
the Bethe equations \eref{eq:BAEsu3} have different types of solutions.
The real roots, called magnons, form the ground state. 
Pairs of complex conjugated roots form bound states of two magnons. 
There are many kinds of bound states.
The so-called string hypothesis was developed to classify such excitations,
for example, for the one-dimensional boson gas with delta interaction
\cite{Yang-Yang}, the Heisenberg chain \cite{TBA}, the
Hubbard model \cite{Takahashi}, the supersymmetric $t-J$
model \cite{Schlottmann} and the Kondo model \cite{Kondo}.
Applying the string hypothesis it is quite standard to derive the TBA equations
\begin{eqnarray}
\fl
\ln (1+\eta _n^{(1)}) &= &
\frac{1}{T}g^{(1)}_n+\sum_{m=1}^{\infty}A_{nm}*\ln(1+{\eta _m^{(1)}}^{-1})
-\sum_{m=1}^{\infty}a_{nm}*\ln(1+{\eta _m^{(2)}}^{-1}),\nonumber\\
\fl
\ln (1+\eta _n^{(2)}) &= &
\frac{1}{T}g^{(2)}_n+\sum_{m=1}^{\infty}A_{nm}*\ln(1+{\eta _m^{(2)}}^{-1})
-\sum_{m=1}^{\infty}a_{nm}*\ln(1+{\eta _m^{(1)}}^{-1}),
\label{eq:TBAe1}
\end{eqnarray}
from \eref{eq:BAEsu3} in a lengthy procedure 
\cite{TBA,Yang-Yang,Schlottmann,TBA2,mix,ying}.
We use the conventional  notation, with
\be
a_n(v)=\frac{1}{2\pi}\frac{n}{n^2/4+v^2}
\ee
and the convolution denoted by $\ast$.
The densities of roots and holes are $\rho^{(k)}_n(v)$ and $\rho^{(k)h}_n(v)$,
for $ k=1, 2$ with 
$\eta_n^{(k)}=\rho^{(k)h}_n(\lambda )/\rho_n^{(k)}(\lambda )$ for
$n=1,2,3,\ldots$. 
The driving terms $g^{(k)}_n$ depend on the choice of eigenbasis. 
We give them below for different configurations, i.e., we will permute the 
basis order to enable the simplest analysis of different regions in the 
parameter space $(H,D,E)$.

We define the so-called {\sl dressed energies} 
$\epsilon^{(k)}_n$ via $\eta _n^{(k)} := \exp(\epsilon^{(k)}_n(\lambda)/T)$ 
with $k=1, 2$.
They play the role of excitation energies, measured from the Fermi energy level for  
each flavour and satisfy Fermi statistics. 
In the limit $T\rightarrow 0$, only the negative part of the dressed energies, 
$\epsilon^{(l)-}:=\min (\epsilon^{(l)},0)$,
contributes to the ground state energy. 
Thus the TBA equations \eref{eq:TBAe1} reduce to two coupled nonlinear integral equations
\begin{eqnarray}
\epsilon^{(1)}&=& g_1  +a_1\ast \epsilon^{(2) -} -a_2\ast \epsilon^{(1) -},
\nonumber\\
\epsilon^{(2)}&=& g_2  +a_1\ast \epsilon^{(1) -} -a_2\ast \epsilon^{(2) -}.
\label{eq:TBAsu3}
\end{eqnarray}
The solution of these equations provides a clear physical picture of the ground state properties. 

\begin{figure}[t]
\begin{center}	
\includegraphics[width=0.70\linewidth]{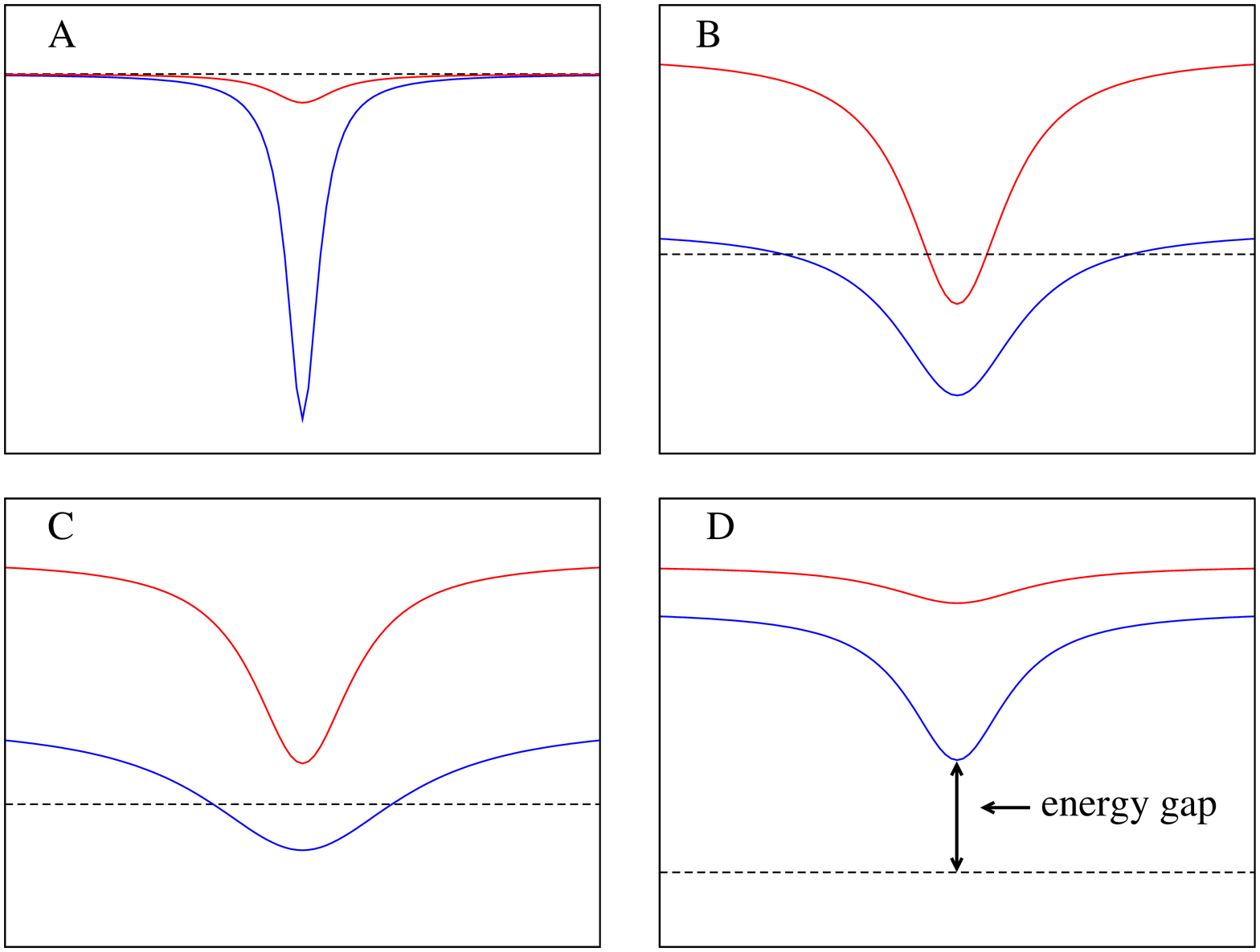}
\end{center}
\caption{Dressed energies for the su(3)-based TBA equations \eref{eq:TBAsu3}.
In each case the horizontal axis is the spectral parameter $v$ with the origin
($v=0$) at the minima. 
The vertical axis carries the dressed energies. 
Due to our basis reordering the upper graph (red) is always
the current second excitation $\epsilon^{(2)}$ while the lower graph
is the first excitation $\epsilon^{(1)}$.  
The Fermi energy for both flavours is always $\epsilon_F=0$, 
denoted by the dotted line. 
Shown are the four typical cases of solutions to \eref{eq:TBAsu3}: 
(A) no chemical potentials, 
(B) mixed state involving all three basis states,
(C) gap in $\epsilon^{(2)}$ and (D) gapped ground state with energy
gap $\Delta$ (see text in \sref{sec:tba}).  }
\label{fig:TBAepscases}
\end{figure}

For typical solutions to \eref{eq:TBAsu3} see \Fref{fig:TBAepscases}. 
To simplify the analysis of \eref{eq:TBAsu3} we separate the calculations for 
regions in parameter space where the basis order changes in accordance with
the energy levels of the chemical potentials.
The assignment $\st[a]$ with $a=1,2,3$ naturally depends on the parameters 
$H,D$ and $E$, which change the energy level of the chemical 
potentials (as, e.g., in \Fref{fig:TBAspeciallevel}).  
The driving terms $g_i$ are then positive or negative in defined regions
\cite{TBA2,mix,ying} allowing for a simple determination of critical fields and gaps. 
The band fillings are controlled by these driving terms. 
In the absence of symmetry-breaking potentials all bands are
completely filled, i.e., their Fermi bundaries are  their value at infinity,
with $\epsilon(\infty)=\epsilon_F$.

\subsubsection{$D>0$, $E>0$}

In this  part of the TBA analysis we discuss the most
general Hamiltonian with both anisotropies, $D>0$ and $E>0$, with an
external parallel magnetic field $H$.  
In later sections we specialize to cases without in-plane anisotropy to 
model certain physical compounds. 
The perpendicular magnetic field term
\eref{eq:Chempotperpendicular} with eigenvalues
\eref{eq:EV.4}-\eref{eq:EV.6} can be treated in the same manner. 
But the in-plane and planar anisotropies play different roles under  
perpendicular and parallel magnetic fields. 

Starting with Hamiltonian \eref{eq:Ham1} without chemical
potentials, i.e. with $\mu=0$, where the model has $su(3)$ symmetry with the
fundamental basis $\st[1]= \st[\!\!+\!\!1]$, $\st[2]= \st[0]$ and $\st[3]=\st[\!\!-\!\!1]$. 
The energy eigenvalue is threefold degenerate.  
The single-ion anisotropy term $D (S^z_j)^2$ breaks this symmetry into a
doublet and a singlet ($su(2)\oplus u(1)$). 
The fundamental basis is still an eigenbasis as the $z$ direction is the 
axis of quantization.
Turning on the external magnetic field, this symmetry is further broken into 
$u(1)\oplus u(1) \oplus u(1)$.
However, the in-plane anisotropy $E$ breaks the $z^2$ symmetry even for  $H=0$. 
The energies split into three levels with respect to the anisotropy and magnetic field. 
The new basis, $\st[1]=a_{-}\st[\!\!-\!\!1]+\st[\!\!+\!\!1]$, $\st[2]=\st[0]$ and
$\st[3]=a_{+}\st[\!\!-\!\!1]+\st[\!\!+\!\!1]$, with $a_{\pm}=[\mu_BgH\pm
\sqrt{(\mu_BgH)^2+E^2}\,]/E$ diagonalizes the potential terms
\eref{eq:Chempotparallel} with eigenvalues \eref{eq:EV.1} to \eref{eq:EV.3}. 
The basis order depends on the external magnetic field if we fix $D$ and $E$. 
Note that still the original singlet state $\st[0]$ remains unchanged.  
The two doublet states $\st[\!\pm\!\!1]$ are rotated in the new 
basis.\footnote{Similar rotations have been considered by Huang and Affleck
in a field theoretic study of the Haldane gap compound NENP \cite{HA}.} 
To display the underlying physics in a clear way it is still convenient to use the 
singlet and doublet terminology, where the ``doublet'' now denotes the rotated 
former doublet states $\st[\!\pm\!\!1]$, even if they are no longer degenerate.

\subsubsection{Case $D>E$}

\begin{figure}[t]
\begin{center}
\vspace{5mm}
\includegraphics[width=1\linewidth]{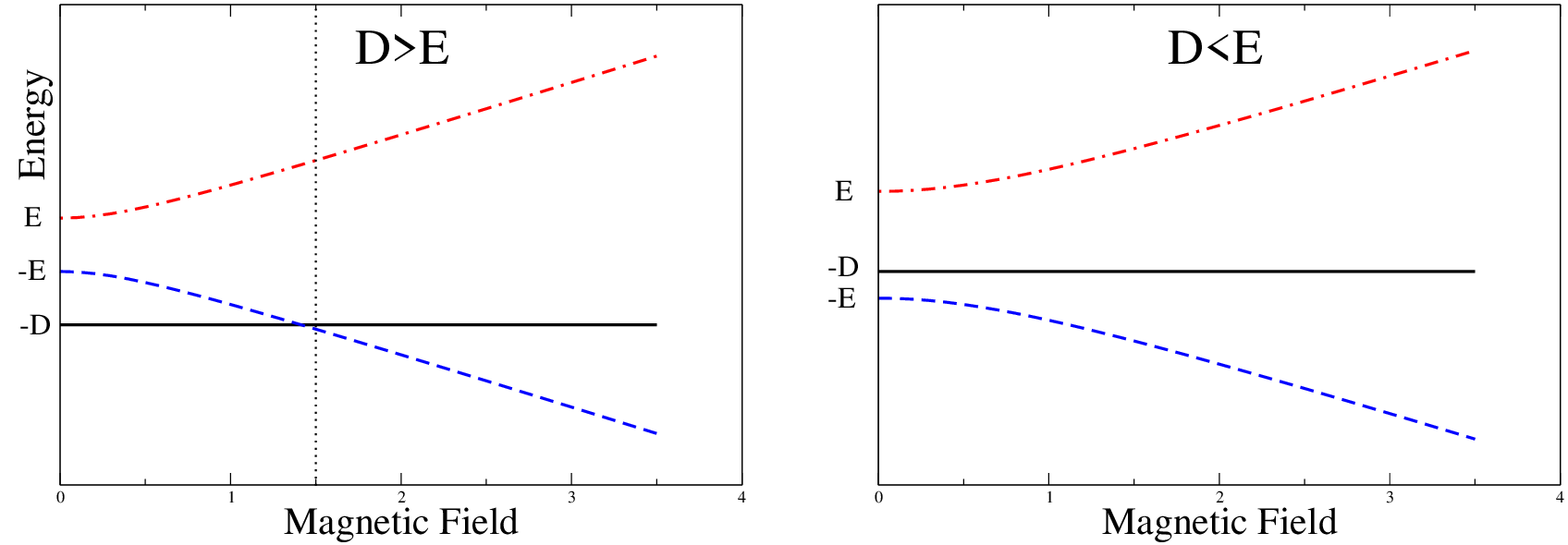}
\caption{Chemical potential energy versus magnetic field for the
parallel field case, equations \eref{eq:EV.1}-\eref{eq:EV.3}, indicating level crossing.
The left figure corresponds to equation \eref{eq:TBAbasisorderingleft} and the right figure
corresponds to equation \eref{eq:TBAbasisorderingright}.
The solid line (black) denotes the singlet $\st[2]$ with energy $\mu_2$,
the dashed line (blue) denotes the doublet state $\st[1]$ with energy $\mu_1$
and the dot-dashed line (red) denotes the doublet state $\st[3]$ with energy $\mu_3$.
In the left figure one level crossing occurs, marked by the dotted vertical line,
while on the right side the basis order for the TBA analysis remains the same for 
all magnetic field. See text and also compare with \Fref{fig:TBAspeciallevel}.
}
\label{fig:TBAsu3elevelscasea}
\end{center}
\end{figure}

In this section we discuss the configuration in which the single-ion
anisotropy $D$ is much stronger than the in-plane anisotropy $E$. 
The applied magnetic field is parallel as in \eref{eq:Chempotparallel}.
In this case the singlet $\st[2]$ is initially energetically
lower than the doublet, as per the left side of \Fref{fig:TBAsu3elevelscasea}.
The components $\st[1]$ and $\st[3]$ further split due to the Zeeman effect. 
In accordance with the energy levels of the chemical potentials, we change the 
basis order as the magnetic field increases.

In the parameter region $H<{\sqrt{D^2-E^2}}/(\mu_B g)$ we have basis order
$\st[2],\st[1],\st[3]$ with TBA driving terms 
\begin{eqnarray}
g_1&=& -\frac{J}{v^2 +\frac{1}{4}} + D - \sqrt{(\mu_BgH)^2 + E^2}, \nonumber\\
g_2&=& 2\sqrt{(\mu_BgH)^2+E^2}. \label{eq:TBAbasisorderingleft}
\end{eqnarray}
Whereas for $H>{\sqrt{D^2-E^2}}/(\mu_B g)$ we have basis order
$\st[1],\st[2],\st[3]$ with driving terms 
\begin{eqnarray}
g_1&=& -\frac{J}{v^2 +\frac{1}{4}} - D + \sqrt{(\mu_BgH)^2 + E^2}, \nonumber\\
g_2&=& D + \sqrt{(\mu_BgH)^2+E^2}. \label{eq:TBAbasisorderingright}
\end{eqnarray}

In this way we see that the parameters $D$, $J$, $E$ and $H$ either raise or lower 
the Fermi surfaces of colours $v^{(1)}$ and $v^{(2)}$.  
A gapped phase occurs if all sites are occupied by the non-magnetic singlet in the
ground state such that it takes a finite energy to excite the doublet. 
In the TBA picture this corresponds to the case where both
{\sl excitation bands} $\epsilon^1$ and $\epsilon^2$ are completely
above the Fermi energy and therefore empty (see configuration D of
\Fref{fig:TBAepscases}).
From the TBA equations in the form \eref{eq:TBAsu3} this can conveniently be 
calculated by finding for which $(H,D,E)$ the gap conditions 
\begin{eqnarray}
\epsilon_1(v=0,H,D,E)>0 \,\,\textrm{\ and\ } \,\,\epsilon_2(v=0,H,D,E)>0
\label{eq:TBAcond1}
\end{eqnarray}
hold.
The restriction $v=0$ could be used because it can be concluded 
from the form of \eref{eq:TBAsu3} that the absolute minima of the 
dressed energies occur at the roots $v=0$.
From \eref{eq:TBAsu3} and \eref{eq:TBAcond1} it can be concluded that
a gapped phase lies in the region 
\be
H<H_{c1}=\sqrt{(D-4J)^2-E^2}/\mu_B g.
\ee
Here $H_{c1}$ is the critical field indicating  a quantum phase transition 
from the nondegenerate singlet to the Luttinger liquid phase. 
The  energy gap is given by 
\be
\Delta =D-4J-\sqrt{(\mu_BgH)^2+E^2}.
\label{eq:gapcond1}
\ee 
This gap is clearly weakened by the in-plane anisotropy as well as the magnetic field.
The necessary condition for this gap to exist can be read off from \eref{eq:gapcond1} 
to be $D>4J+E$.  
In the dressed energy picture this manifests itself by the point at which the lowest
dressed energy $\epsilon ^{(1)}$ just tips the Fermi energy $\epsilon_F=0$
at its minimum, i.e. the transition D $\to$ C in \Fref{fig:TBAepscases}.  
Note that due to our basis-reordering procedure the dressed energies 
always satisfy $\epsilon^{(2)} (v)\geq \epsilon^{(1)}(v)$.  
Thus once $H>H_{c1}$ the component $\st[1]$ with
eigenvalue \eref{eq:EV.1} becomes involved in the ground state.  
At the critical point $H_{c1}$ the phase transition is not of 
Pokrovsky-Talapov type, due to the mixture of the former doublet states
$\st[S^z=\pm 1]$ in the state $\st[1]$ with a
field-dependent magnetic moment.

The magnetization increases as the magnetic field increases while 
more of component $\st[1]$ becomes involved in the ground state.
As usually defined in the TBA approach, the magnetization
of an eigenstate of the system is
\be M(H)= \sum_{j=1}^{3} N_{\st[j]}  m_{\st[j]},
\label{eq:deftbamag}
\ee
where the numbers $N_{\st[j]}$ denote the total number of state $\st[j]$ with magnetic moment
\be
m_{\st[j]}= -\frac{\partial}{\partial H} \mu_{j}(H). 
\label{eq:deftbamag2}
\ee
Here $\mu_{j}$ are the eigenvalues of the chemical potentials.
The occupation numbers $N_{\st[j]}, j=1,2,3$ are  related to the Bethe quantum
numbers $M_1$ and $M_2$ via \eref{eq:BANMrelation}, where $M_1$ and
$M_2$ are given by
\be
M_1 = \int \rho^{(1)}(v) \rmd v \qquad \textrm{ and }\qquad M_2 = \int
\rho^{(2)}(v)\rmd v.  
\ee 

\begin{figure}
\begin{center}
\vspace{5mm}
\includegraphics[width=.60\linewidth]{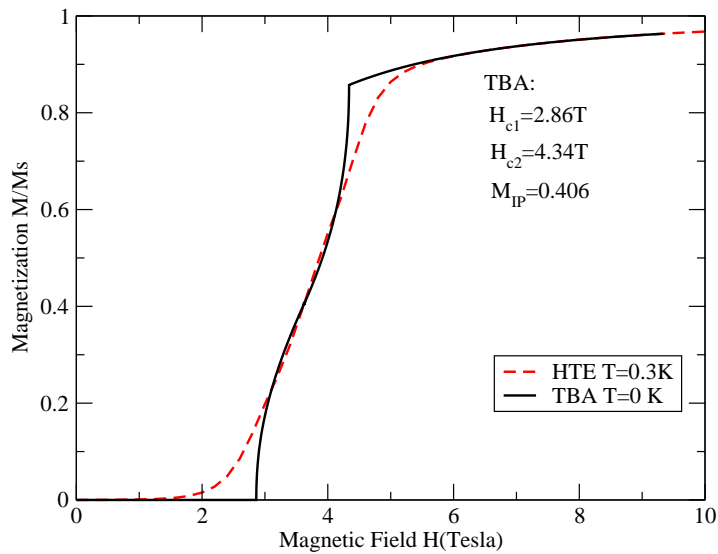}
\caption{Magnetization versus magnetic field $H$ in units of
saturation magnetization for the Hamiltonian \eref{eq:Ham1} with
$J=0.2$ K, $D=6$ K, $E=3.5$ K and $g=2.0$ with parallel magnetic field. 
The solid and dashed lines show the magnetization derived from
TBA and HTE at $T=0$ and $T=0.3$ K, respectively. 
The TBA magnetization curve indicates quantum phase transitions in
the vicinity of $H_{c1}$ and $H_{c2}$ differing from the square root
field-dependent critical behaviour. 
Their middle point has magnetization $M_{{\rm IP}} \approx 0.406$ 
rather than $M_{{\rm IP}}=0.5$ in the case $E=0$. 
Overall the TBA calculation is consistent with the HTE result.}
\label{fig:SP1SZ}
\end{center}
\end{figure}

From the driving terms in \eref{eq:TBAbasisorderingright} 
we see that once the field increases beyond the second critical point 
\be
H_{c2}=\frac{\sqrt{(D+4J)^2-E^2}}{\mu_B g}, 
\ee 
the ground state consists entirely of the $\st[1]$ state.
In this special case $M_1=M_2=0$ holds, and therefore the reference state $\st[1]$
becomes a true physical state with (normalized) magnetization 
\begin{eqnarray}
M_{H>H_{c2}}(H) = \frac{\mu_BgH}{\sqrt{E^2+ (\mu_BgH)^2}}.
\label{eq:Mag1}
\end{eqnarray}
We thus see that the magnetization in the pure $\st[1]$ state gradually 
approaches the saturation magnetization $M_s=1$ as the contribution to
the magnetization from the original doublet state $\st[S^z\!\!=\!\!-1]$ in the
mixed state $\st[1]$ tends to zero, whereas the state $\st[S^z\!\!=\!\!-1]$ in
the mixed state $\st[3]$ becomes dominated.
The probabilities of the components $\st[1]$ and $\st[2]$ are equal  
at the inflection point $H=\sqrt{D^2-E^2}/\mu_Bg$, where the magnetization
$M_{{\rm IP}}=\frac{1}{2}\sqrt{1-(E/D)^2}$.
Actually, the magnetization between the critical fields can be
obtained by numerically solving the TBA equations \eref{eq:TBAsu3}.
This reduces to solving one integral equation instead of two coupled
integral equations because the second level dressed energy is gapful
in many cases (i.e. $\epsilon^{(2)}(v)>0$).
This numerically convenient case occurs when the general part B 
of \Fref{fig:TBAepscases} reduces to part C.
Indeed, in \Fref{fig:SP1SZ}, these novel phase transitions are observed in the
magnetization derived from the TBA equations \eref{eq:TBAsu3} and HTE
free energy given by \eref{eq:freeenergyHTE}.
As the spin-spin exchange interaction $J$ decreases, the 
magnetization increases steeply in the vicinity of the critical point $H_{c1}$. 
If $J=0$, i.e., for the case of independent spins, the critical points 
$H_{c1}$ and $H_{c2}$ merge into one point, at which a discontinuity in the 
magnetization of the ground state occurs. 
This is in agreement with the trivial exact solution of
a single-spin system with energy levels \eref{eq:EV.1}-\eref{eq:EV.3}.

\subsubsection{Case $D\leq 4J+E$} 

If the single-ion anisotropy $D$ lies in the region  $E-4J<D\leq E+4J$,
the energy levels of the chemical potentials \eref{eq:Chempotparallel}
are as shown on the right side of \Fref{fig:TBAsu3elevelscasea}. 
The gapped phase does not exist in this regime. 
There are two competing components in the ground state, namely $\st[1]$ and $\st[2]$.
In this case the basis order is $\st[1],\st[2],\st[3]$ and the driving terms
\begin{eqnarray}
g_1&=& -\frac{J}{v^2+ \frac{1}{4}} -D + \sqrt{E^2+ (\mu_B g H)^2} \nonumber\\
g_2  &= &D +\sqrt{E^2+ (\mu_B g H)^2}
\end{eqnarray}
apply throughout the whole range of the magnetic field.  
For very large magnetic field, i.e., $H>H_c$,  where 
$H_c=\sqrt{(4 J +D)^2 - E^2}/\mu_B g$, the ground state
consists purely of the doublet state $\st[1]$. 
Again, in the dressed energy picture, $H_c$ is the point at which the first
excitation band, here given by the singlet state $\st[2]$, touches the
Fermi energy $\epsilon_F=0$ at its minimum, as per the transition 
D $ \to$ C in \Fref{fig:TBAepscases}. 
To the right, i.e., for higher magnetic fields, both excitation bands 
$\epsilon^{(1)}$ and $\epsilon^{(2)}$ are completely gapped 
(part D of \Fref{fig:TBAepscases}). However, in the case $D<E-4J$, 
all three states are involved in the ground state in a gapless phase. 

\subsubsection{Perpendicular magnetic field} 

If a perpendicular magnetic field is applied to Hamiltonian
\eref{eq:Ham1}, similar quantum phase transitions might occur as  long as the in-plane anisotropy is much stronger than the planar anisotropy. Namely, if $E>D+4J$, a gapped phase lies in the region $H<H_{c1}$, where the critical field 

 \be H_{c1}=\frac{\sqrt{2(E-2J)(E-4J-D)}}{\mu_B g}.  
\ee 
The magnetization increases almost linearly as the magnetic field
increases above $H_{c1}$. A quantum phase transition
occurs at the critical point $H_{c2}$, where 
\be
H_{c2}=\frac{\sqrt{2(E+2J)(E+4J-D)}}{\mu_B g}.  
\ee
The magnetization tends to saturation as the field becomes stronger
than $H_{c2}$. However, if $E<D+4J$, there is no gapped phase even in
the absence of the magnetic field.

\subsubsection{Special case 1: Absence of in-plane anisotropy ($E=0$)}
\label{sc1}

Now consider Hamiltonian \eref{eq:Ham1}
with chemical potentials \eref{eq:Chempotparallel} for the special case $E=0$. 
This will be used to analyze the compound  NENC in \sref{sec:NENC}, 
which is believed to have negligible in-plane anisotropy.
The fundamental basis, $\st[1]= \st[\!\!+\!\!1]$, $\st[2]= \st[0]$ and $\st[3]=\st[\!\!-\!\!1]$, 
diagonalizes the chemical potential terms with eigenvalues
\begin{eqnarray}
\mu_1&=-\mu_B g H ,\nonumber\\
\mu_2&= -D,\nonumber\\
\mu_3&= \mu_B g H .
\label{eq:chempot2}
\end{eqnarray}
Taking $E=0$ in the driving terms in the different regions, we find that
the ground state in the zero temperature limit has a gap if the single-ion
anisotropy $D > 4J$.  
The singlet $\st[2]$ ground state is separated from
the lowest spin excitation by an energy gap $\Delta =D-4J$. 
This energy gap is decreased by the external magnetic field $H$. 
The pure singlet ground state breaks down at the critical point 
$H_{c1}=(D-4J)/\mu_Bg_s$.
The magnetization almost linearly increases with increasing magnetic field 
due to the magnon excitation. 
We discuss this further via the HTE approach in \sref{HTE}. 
The ground state is fully polarized once the magnetic field 
increases beyond the second critical point $H_{c2}=( D+4J)/(\mu_Bg)$, 
i.e., in the $M\!=\!M_s$ saturation plateau region. 
A gapped phase can only exist for anisotropy values satisfying the `strong anisotropy' 
condition $D>4J$. 
As shown in \cite{TBA2} the magnetization in the
vicinity of the critical fields $H_{c1}$ and $H_{c2}$ depends on the
square root of the field, indicating a Pokrovsky-Talapov type transition.
In this regime, the anisotropy effects overwhelm the biquadratic
spin-spin interaction contribution and open a gapped phase in the ground state.

\subsubsection{Special case 2: Trivalent orbital splitting}

Here we consider a further specialization of the previous
case \eref{eq:Chempotparallel} which can be mapped onto it via the
identification
\be 
(H',D',E')=(H+2J_1,D-2J_1,0) ,
\ee
where the primed constants are those apearing in the general 
model \eref{eq:Chempotparallel}.
We have not found an experimental spin-compound realization of
this particular Hamiltonian so far. 
Nevertheless, we give the TBA analysis here, as it has new and interesting physics
in its own right, as it incorporates a physical mechanism to open more unusal gapped
states and magnetization plateaux.

\begin{figure}[t]
\begin{center}
\vspace{5mm}
\includegraphics[width=\linewidth]{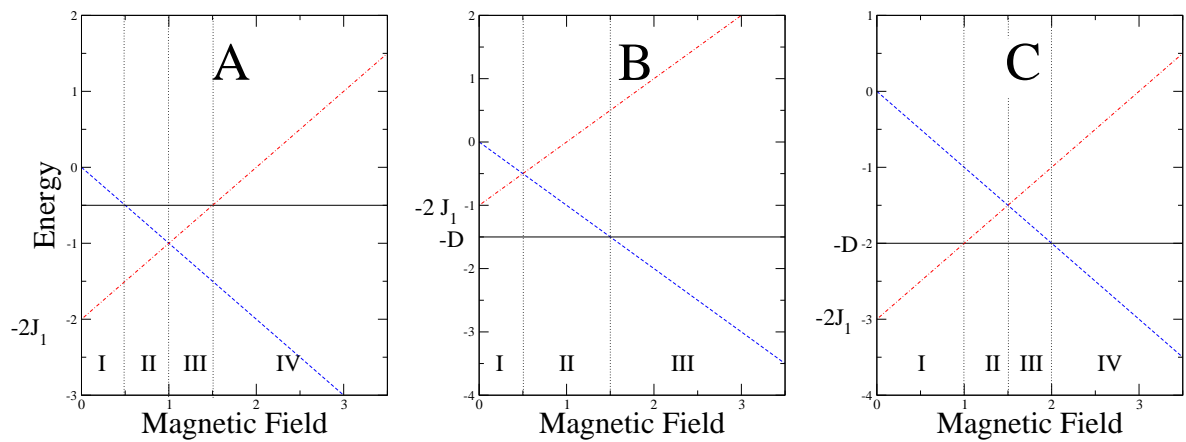}
\caption{Energy levels versus magnetic field for eigenvalues 
\eref{eq:Evalspec} of the operator \eref{eq:SPC2}.
The solid line (black) denotes the singlet $\st[0]$, the dashed line (blue)
denotes the doublet $\st[{+}1]$ and the dash-dotted line (red) 
denotes the other doublet state $\st[{-}1]$.
The vertical dotted lines and the roman numerals mark the change of basis 
order in the TBA analysis (see text). 
Shown are the three typical choices of coupling constants. 
For their connection to energy gaps and magnetization plateau see the text 
and also \Fref{fig:HTEfig1} and \Fref{FIG2}.}
\label{fig:TBAspeciallevel}
\end{center}
\end{figure}
 
If we consider the orbital degrees in the Hamiltonian,
trivalent orbital splitting, as well as interatomic effects, 
may result in an alignment of the internal magnetic field
to prefer one component of the spin doublet. 
To represent this physical effect we add a projection operator
to the chemical potentials, so that
\be
{\cal H}_{i,\textrm{\tiny chem.pot}}= D \sum_{j=1}^L({S}_j^z)^2 - J_1 
\sum_{j=1}^L{S}^z_j ({S}^z_j -1)
- \mu_B g H \sum_{j=1}^L{S}^z_j.
\label{eq:SPC2}
\ee
This additional term gives rise to a rich phase diagram by shifting the energy
levels, as seen in the difference between \Fref{fig:TBAsu3elevelscasea}
and \Fref{fig:TBAspeciallevel}.
The on-site energy eigenvalues are now given by
\begin{eqnarray}
\mu_{\st[+1]}=-\mu_B g H,
\nonumber\\
\mu_{\st[0]}= -D,
\label{eq:Evalspec}\\
\mu_{\st[-1]}=\mu_B g H - 2 J_1, \nonumber
\end{eqnarray}
where the fundamental basis, $\st[1]= \st[\!\!+\!\!1]$, $\st[2]= \st[0]$ and
$\st[3]= \st[\!\!-\!\!1]$, diagonalizes the chemical potential term.

In general, the magnetization curves for this case can have up to $3$
gapped phases, as shown in part (B) of \Fref{fig:HTEfig1}.
Note that all magnetization plots in this figure have been obtained
by the HTE method for relatively low temperatures and are therefore
only similar to the zero-temperature magnetization. 
Numerical solution of the TBA equations gives slightly `sharper' contours. 
Variations in the parameter set ($J$, $J_1$, $D$) can open
or close the different gapped phases. 
After calculating the whole phase diagram via the TBA, 
analogously as done in Refs~ \cite{TBA2,mix,ying}, we 
show the physically interesting choices of coupling constants,
corresponding to the four pictures in \Fref{fig:HTEfig1}. 
Here we have restricted our analysis to small exchange coupling values $J<D/4$.  
We now consider the model in the region $J_1 < D < J_1 + 4 J$, 
corresponding to configuration A of \Fref{fig:TBAspeciallevel}.
Proceeding  with the basis orders
\begin{eqnarray}
(\st[3],\st[2],\st[1])\rightarrow (\st[3],\st[1],\st[2])\rightarrow 
(\st[1],\st[3],\st[2])\rightarrow (\st[1],\st[2],\st[3]),
\label{eq:basisorder1}
\end{eqnarray}
we analyze the ground state properties from the TBA equations
\eref{eq:TBAsu3} with different driving terms. 
The technique is similar to the previous section and we 
omit the details of the driving terms.
It is interesting that a degenerate gapped phase starts from zero field.  
The ground state consists of the doublet state $\st[3]$ with spin $S^z\!=\!-1$. 
This state is separated from the lowest magnon excitation by the energy gap 
\be 
\Delta
=2J_1-D-4J. 
\ee 
The magnetic field lifts the energy level of the state $\st[3]$\
and the gap vanishes when the magnetic field exceeds the first critical field
\be
H>H_{c1}= \frac{2J_1-D-4J}{\mu_Bg}.
\ee
The  singlet state and the other doublet state  $\st[1]$  subsequently become involved 
in the ground state. 
Thus the magnetization increases almost linearly as the magnetic field increases. 
It is worth noting that the singlet state is suppressed by the strong potential $J_1$. 
The ground state is fully-polarized beyond the second critical magnetic field 
\be
H>H_{c2}=\frac{D+4J}{\mu_Bg}.
\ee
This phase diagram is shown in part (A) of \Fref{fig:HTEfig1}.

In the region $J_1+4 J <D<2 J_1 -4 J$, we find the situation as
depicted in configuration (C) of \Fref{fig:TBAspeciallevel},
where the model exhibits three different gapped phases due to existence
of a strong single-ion anisotropy. 
Here the basis orders become 
\begin{eqnarray}
(\st[3],\st[2],\st[1])\rightarrow (\st[2],\st[3],\st[1])\rightarrow 
(\st[2],\st[1],\st[3])\rightarrow (\st[1],\st[2],\st[3]).
\label{eq:basisorder2}
\end{eqnarray}
The plateau with total spin $S^z=-1$ starts at zero field. 
The energy level of this state is increased by the external magnetic field, 
so that the spin singlet completely displaces the doublet from the ground state
 when the field exceeds $H>H_{c2}$. 
This field-induced nonmagnetic mid-plateau opens 
with the anisotropy parameter $D$. 
However, it can be closed for certain large values of $J$.  
The critical fields predicted by the TBA analysis are given by
\begin{eqnarray}
H_{c1} &= \frac{2J_1-D-4J}{\mu_Bg},\qquad\qquad  & 
H_{c2}= \frac{2J_1-D+4J}{\mu_Bg},\nonumber\\ & & \label{CF}\\
H_{c3} &= \frac{D-4J}{\mu_Bg},  &H_{c4}= \frac{D+4J}{\mu_Bg}. 
\nonumber
\end{eqnarray}
Further physical interpretation of these critical fields is given in section \ref{HTE} via the
HTE approach (see part (B) in \Fref{fig:HTEfig1}). 
If the single-iron anisotropy becomes larger, such that  
$2J_1-4J\!<\!D\!<\!2J_1+4J$, the gapped spin $S^z=-1$ state is weakened by the 
singlet, whose energy level decreases. 
Thus the doublet magnetic ground state is gapless as long as the field is below $H_{c2}$.
In this case the mid-plateau can still survive (see part (C) in \Fref{fig:HTEfig1}). 
Except for the absence of the first critical field $H_{c1}$,  
the other critical fields in \eref{CF} remain  valid for this case. 
However, if $D$ is large enough a nondegenerate singlet state can start from zero field.  
This scenario is  depicted in part {(B)} of \Fref{fig:TBAspeciallevel}. 
In this case, the anisotropy $J_1$ is suppressed. 
The critical fields are the same as in \sref{sc1}.

\subsection{HTE analysis: properties at high temperatures}
\label{HTE}
\label{sec:hte}

We begin by briefly reviewing the idea of the HTE method, which we use
to predict physical properties of the spin chain compounds in the next
section.

The Bethe Ansatz as described in section 2.1 is based on a 
row-to-row transfer matrix, which has the disadvantage that its largest eigenvalue 
may be degenerate.
Moreover, the TBA equations suffer from infinitely many nonlinear integral equations (NLIE). 
On the other hand, the Quantum Transfer Matrix (QTM) \cite{QTMrefs} considers the 
transfer matrix on a square lattice with an inhomogeneity  parameter value 
$u_N = -\beta J/N$ ($N$ is the Trotter number). 
The largest eigenvalue is non-degenerate \cite{QTMproof} and the free energy is 
determined only by  the largest eigenvalue of the QTM. 
The advantage of the QTM formalism is that the problem of deriving the largest 
eigenvalue reduces to solving finitely many NLIE. 
The nested Bethe equations take a slightly different form than in 
the row-to-row case.

For the model at hand, the eigenvalue of the QTM is given by
\begin{eqnarray}
T^{(1)}_1(v,\left\{v^{(a)}_i\right\})
&=%
e^{-\beta \mu_1}\ 
\phi _-(v-\mathrm{i})\phi_+(v)
&\frac{Q_1(v\!+\!\frac12\mathrm{i})}{Q_1(v\!-\!\frac12\mathrm{i})}
\nonumber\\
&+e^{-\beta \mu_2} \ 
\phi _-(v)\phi _+(v)
&\frac{Q_1(v\!-\!\frac32\mathrm{i})}{Q_1(v\!-\!\frac12\mathrm{i})}\
\frac{Q_2(v)\ \ }{Q_2(v\!-\!\mathrm{i})}
\nonumber\\
&+
e^{-\beta \mu_3}\ 
\phi _-(v)\phi _+(v+\mathrm{i})
&\frac{Q_2(v\!-\!2\mathrm{i})}{Q_2(v\!-\!\mathrm{i})\ }.
\label{EQTM1}
\end{eqnarray}
Here we follow notation from Ref.~\cite{ZT}, with
$\phi _{\pm}(v)\!=\!(v\pm  \mathrm{i}u_N)^{\frac{N}{2}}$ and
$Q_a(v)=\prod_{i=1}^{M^{(a)}}(v\!-\!v_i^{(a)})$ for $a=1,2$. 
The Bethe roots $v_i^{(a)}$ are solutions of the QTM Bethe equations
\begin{eqnarray}
& &\left(\frac{v_j^{(1)}-\mathrm{i}u_N+\frac12\mathrm{i}}
{v_j^{(1)}-\mathrm{i}u_N-\frac12\mathrm{i}}\right)^{\frac{N}{2}}
={\rm e}^{-\frac{\mu_1-\mu_2}{T}}\prod^{M_1}_{\stackrel{\scriptstyle l=1}{l\neq j}}
\frac{v_j^{(1)}-v_l^{(1)}+\mathrm{i}}{v_j^{(1)}-v_l^{(1)}-\mathrm{i}}
\prod_{l=1}^{M_{2}}\frac{v_j^{(1)}-v_l^{(2)}-\frac12\mathrm{i}}
{v_j^{(1)}-v_l^{(2)}+\frac12\mathrm{i}},\nonumber \\
& &
\left(\frac{v_k^{(2)}+\mathrm{i}u_N+2\mathrm{i}}
{v_k^{(2)}+\mathrm{i}u_N+\mathrm{i}}\right)^{\frac{N}{2}}=
{\rm e}^{-\frac{\mu_2-\mu_3}{T}}\prod_{i=1}^{M_{1}}\frac{v_k^{(2)}-v_i^{(1)}-\frac12\mathrm{i}}
{v_k^{(2)}-v_i^{(1)}+\frac12\mathrm{i}}
\prod^{M_2}_{\stackrel{\scriptstyle l=1}{l\neq k}}
\frac{v_k^{(2)}-v_l^{(2)}+\mathrm{i}}{v_k^{(2)}-v_l^{(2)}-\mathrm{i}}.
\label{eq:qBAEsu3}
\end{eqnarray}
In the above $j=1,\ldots, M_1$ and $k=1,\ldots, M_2$.

The eigenvalue $T_1^{(1)}(v)$ is embedded in a family of eigenvalues $T^{(a)}_m(v)$ 
for related models via the fusion hierarchy \cite{KunibaFusion}. 
These eigenvalues are connected to each other by the functional relation
\eref{eq:TSystem} known as the T-system. 
The indices of $T^{(a)}_m$ denote that the auxiliary space has fusion
type $(a,m)$. It carries the $m$-fold symmetric tensor of the $a$-th
fundamental representation of the $su(3)$ Lie algebra.
The relations can be conveniently denoted graphically by Young 
tableaux, where the simple case only deals with 
rectangular shaped blocks (see Refs~\cite{ZT} and \cite{KunibaFusion}).
The fused eigenvalues are explicitly given by
\begin{equation}
T^{(a)}_m=\sum_{d_{j,k}}\prod_{j=1}^{a}\prod_{k=1}^mz(d_{j,k};v-\case12\mathrm{i}(a-m-2j+2k)),
\end{equation}
where the summation is taken over the ``Young tableaux'' parameter 
$d_{j,k} \in  \left\{1,2,3\right\}$ such that $d_{j,k} < d_{j+1,k}$ and $d_{j,k} \leq d_{j,k+1}$, 
which form an $a\times m$ rectangular tableaux. 
After a renormalization the functional relation takes the simple form
\begin{equation}
\tilde{T}^{(a)}_m(v+\case12{\mathrm{i}})\tilde{T}^{(a)}_m(v-\case12{\mathrm{i}})=
\tilde{T}^{(a)}_{m+1}(v)\tilde{T}^{(a)}_{m-1}(v)+\tilde{T}^{(a-1)}_{m}(v)\tilde{T}^{(a+1)}_{m}(v).
\label{eq:TSystem}
\end{equation}
in the normalized  functions $\tilde{T}^{(a)}_m$ \cite{ZT}.

By exploiting knowledge of analyticity properties, and numerically backed 
assumptions about the largest eigenvalue and the locations of its zeroes, 
this T-system can be transformed into a set of coupled NLIE.
These NLIE had first been discussed for the spin-$\frac{1}{2}$ chain 
($su(2)$-XXX model) in \cite{HTE2}. 
The general $su(n)$ case was derived by Tsuboi \cite{ZT} in full detail.
For the integrable spin-$1$ chain ($su(3)$-case) considered here the coupled
NLIE are
\begin{eqnarray}
\fl
T_1^{\textrm{\tiny (0)}}(v) &=\exp
\left({\frac{J}{T} \frac{1}{v^2 + \frac{1}{4}}}\right),\nonumber \\
\fl
T_1^{(1)}(v)& = Q_1^{(1)}
{{+}}\oint_{c_1^1} \frac{\textrm{d}y}{2\pi\mathrm{i}}
\frac{T_1^\textrm{\tiny(0)}(y{{+}}\beta_1^\textrm{\tiny(1)}{ 
-}\frac12\mathrm{i})  T_1^\textrm{\tiny(2)}(y{{+}}\beta_1^\textrm{\tiny(1)}{ 
-}\frac12\mathrm{i})
}{(v{ -} y  { -} 
\beta_1^\textrm{\tiny(1)})T_1^\textrm{\tiny(1)}(y{{+}}\beta_1^\textrm{\tiny(1)}{
 -}\mathrm{i})}
{{+}}\oint_{c_2^1}\left\{
{\mathrm{i} \shortrightarrow { -} \mathrm{i} \atop  {\beta_1^\textrm{\tiny(1)} \shortrightarrow 
{{-}} \beta_1^\textrm{\tiny(1)}}}
\right\},
\nonumber\\
\fl
T_1^{(2)}(v) & = Q_2^{(1)}
+ \oint_{c_1^2}\! \frac{\textrm{d}y}{2\pi\mathrm{i}}
\frac{T_1^\textrm{\tiny(1)}(y{\scriptstyle{+}}\beta_1^\textrm{\tiny(2)}{
-}\frac12\mathrm{i}) 
T_1^\textrm{\tiny(3)}(y{{+}}\beta_1^\textrm{\tiny(2)}{{-}}\frac12\mathrm{i})
}{(v{{-}} y  {{-}} 
\beta_1^\textrm{\tiny(1)})T_1^\textrm{\tiny(2)}(y{{+}}\beta_1^\textrm{\tiny(2)}{
 -}\mathrm{i})}
{+} \oint_{c_2^2}
\left\{ {
\mathrm{i} \shortrightarrow - \mathrm{i} \atop  
{\beta_1^\textrm{\tiny(2)}  \shortrightarrow  { - \beta_1^\textrm{\tiny(2)}}}
}\right\},
\nonumber\\
\fl
T_1^{(3)}(v)& = \exp\left( (\mu_1+\mu_2+\mu_3)/T \right)=\textrm{const.} 
\label{eq:NLIE3}
\end{eqnarray}
Here the contours $c_i^j$ are counterclockwise closed loops around zero, 
omitting certain special points \cite{ZT}.

The first remarkable property of these new NLIE's is that there are only finitely 
many of them, as the $n$-th eigenvalue in this hierarchy is constant
in terms of the spectral parameter $v$, in contrast 
to the finite temperature TBA considered in \sref{sec:tba}.
The second useful property is the fact that these NLIE's can 
be solved by an expansion in the small parameter $J/T$,
for which the coefficients can be obtained {\sl recursively}.
This distinguishes the HTE method from other NLIE approaches, 
e.g., Ref.~\cite{QTMproof}, where the NLIE has to be solved numerically.
 
The result for the free energy of the $su(3)$ case (already stated in Ref.~\cite{ZT}) is 
\begin{equation}
-\frac{1}{T}f(T,H)=\ln Q^{(1)}_1+C^1_{1,0}\left(\frac{J}{T}\right)
+C^1_{2,0}\left(\frac{J}{T}\right)^2+C^1_{3,0}\left(\frac{J}{T}\right)^3+\cdots
\label{eq:freeenergyHTE}
\end{equation}
where the first few coefficients are given by
\begin{eqnarray}
c_{1,0}^{(1)}&=2 \frac{Q^{(2)}_1}{{Q^{(1)}_1}^2}, \nonumber\\
c_{2,0}^{(1)}&=3 \frac{Q^{(2)}_1}{{Q^{(1)}_1}^2}- 6 \frac{{Q^{(2)}_1}^2}{{Q^{(1)}_1}^4} 
+3 \frac{{Q^{(3)}_1}}{{Q^{(1)}_1}^3},\nonumber\\
c_{3,0}^{(1)}&= \frac{10}{3}\frac{Q^{(2)}_1}{{Q^{(1)}_1}^2}
-18\frac{{Q^{(2)}_1}^2}{{Q^{(1)}_1}^4}	 
+\frac{80}{3}\frac{{Q^{(2)}_1}^3}{{Q^{(1)}_1}^6}
+8\frac{{Q^{(3)}_1}}{{Q^{(1)}_1}^3}
-24\frac{{Q^{(2)}_1}{Q^{(3)}_1}}{{Q^{(1)}_1}^5},
\nonumber 
\end{eqnarray}
in terms of the su(3) Q-system 
\begin{eqnarray}
Q_1^{(1)} =& e^{-\beta \mu_1}+e^{-\beta \mu_2}+e^{-\beta \mu_3},\nonumber \\
Q_1^{(2)} =& e^{-\beta\mu_1-\beta\mu_2}+e^{-\beta\mu_1-\beta \mu_3}+e^{-\beta \mu_2-\beta\mu_3},
\label{eq:QSystem}\\
Q_1^{(3)} =& e^{-\beta\mu_1 -\beta \mu_2 -\beta \mu_3}.\nonumber
\end{eqnarray}

We find that considering only terms up to third order is sufficient for the analysis of 
real compounds.
This is different from other types of high temperature series expansion \cite{HTserie}.
The HTE coefficients are functions of the
chemical potentials $\mu_i$, i.e., they depend on the external parameters
such as the coupling strength and magnetic field.  
The thermodynamic behaviour can be directly obtained from the free energy
\eref{eq:freeenergyHTE} via the standard relations
\begin{equation}
\begin{array}{lll}
\fl
M= {\scriptstyle{-}}
\left(
\frac{\partial}{\partial H} f(T,H)
\right)_{\!\!T}, \,\,&
\chi ={\scriptstyle{-}}
\left(
\frac{\partial^2}{\partial H^2} f(T,H)
\right)_{\!\!T},\,\,&
C= {\scriptstyle{-}} T\! 
\left(
\frac{\partial^2}{\partial T^2} f(T,H)
\right)_{\!\!H}, \nonumber 
\end{array}
\end{equation}
for the magnetization, magnetic susceptibility and magnetic specific heat.
We apply these results to several spin compounds in \sref{sec:exp}. 
Now the {\sl small parameter} of the expansion is $J/T$, i.e., the expansion is 
expected to work well for spin chain materials with relatively weak intrachain coupling.

Before proceding, we note that in the above theory, the largest QTM eigenvalue is only
 fixed up to a (positive) constant. 
From the derivation of the HTE it can be seen that the free energy expansion
\begin{equation}
f_{HTE} (T,H)= c_0 + c_1 \frac{J}{T}
+c_2 \left(\frac{J}{T}\right)^2
+c_3 \left(\frac{J}{T}\right)^3 +...
\end{equation}
is only fixed up to a multiplicative constant, i.e.
\begin{eqnarray}
f_{{\rm phys.units}}&=& \gamma_{{\rm conv.}}  f_{HTE} (T,H)\nonumber\\
&=&\gamma_{{\rm conv.}} \left[c_0 + c_1 \frac{J}{T}
+c_2 \left(\frac{J}{T}\right)^2
+c_3 \left(\frac{J}{T}\right)^3 +...\right].
\label{eq:HTEfree-c}
\end{eqnarray}
Physically, this amounts to the fact that in the theory we don't specify the unit of energy, i.e. there is no analogue of 
a `characteristic length'. Mathematically it can be traced back to the multiplicativity of the underlying $R$-matrix of the 
integrable model. 
In practice this factor depends on the choice of units for the physical free energy.  

Another remaining problem is the apparent widespread use of non-standard non-SI units in the experimental community 
for measurements on real compounds which sometimes even depend on 
compound specifics such as molecular mass. 
In eq.(\ref{eq:HTEfree-c}) the unknown constant $\gamma_{{\rm conv.}}$ is a prefactor, i.e., the same in 
all orders of the expansion. 
The expansion is generated for interacting spins from the Bethe Ansatz. 
For zero order,  the non-interacting case $J=0$,  it reduces to the exact trivial solution for $L$ 
independent spins (here, e.g., for a 2-state spin-1/2 system):
\begin{equation}
f_{{\rm 0-th\, order}} = \gamma_{{\rm  conv.}} c_0
\equiv -k_B T \ln \left(    e^{\frac{\mu_1}{k_B T}}+e^{\frac{\mu_2}{k_B T}}     \right).
\end{equation}
The second equality only holds for the special point $J=0$, but as $\gamma_{{\rm  conv.}}$ is a constant
it can be extended in eq.(\ref{eq:HTEfree-c}) to all $J>0$.

Once the free energy is thus known in SI units, say [J], it is trivial to obtain all other properties dervied from it in
SI units via
\begin{equation}
M_{{\rm  phys. units}}=-\frac{\partial}{ \partial H} f_{{\rm phys. units}}.
\end{equation}
The important point here is that HTE can make predictions in physical
units, even if the original T-system is only valid up to a constant. 
Thus all conversion factors can in theory be calculated
using the additional  information from the non-interacting spin system. 
In addition, for real compounds the parameters $D$, $E$ and $J$ are given in Kelvin. 
For convenience, we omit in this paper the Boltzmann constant $k_{B}$ in the conventional
notations ${D}/{k_B}$, ${E}/{k_B}$, ${J}/{k_B}$.

We found different  spin-$1$ compounds have the same conversion factor
for the  susceptibility. 
In this way if we normalise the magnetic properties, such
as the magnetization ($M/M_s$), the conversion constant does not play any
role in fitting the experimental data. 
On the other hand, the conversion constant for the specific heat is material dependent 
due to the choice of units in a given experiment. 
For convenience we state these factors in the relevant figure
caption each time they are used.

\subsubsection{Special case: Trivalent orbital splitting}

In this section we consider the special  Hamiltonian \eref{eq:SPC2}, 
already studied via the TBA method. 
Diagonalizing the Hamiltonian we find the chemical potential terms \eref{eq:Evalspec}
\begin{eqnarray}
\mu_1&=-\mu_BgH,\nonumber\\
\mu_2&=-D, \label{eq:Evalssimplemodel}\\
\mu_3&=\mu_BgH-2J_1. \nonumber
\end{eqnarray}

\begin{figure}[t]
\begin{center}
\vspace{10mm}
\includegraphics[width=.70\linewidth]{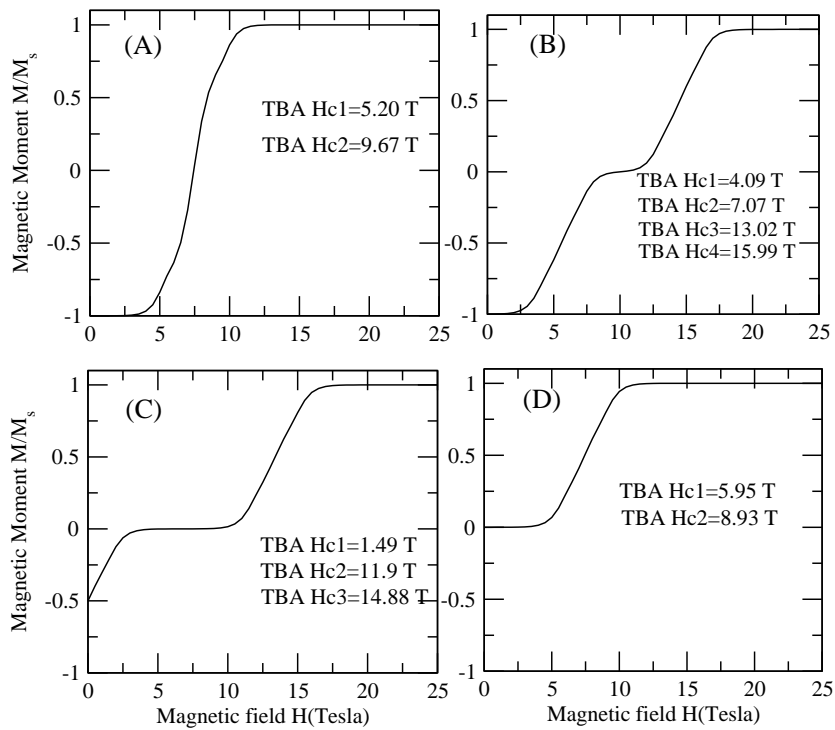}
\end{center}
\caption{ 
Magnetization versus magnetic field at $T=0.8$K for the   
Hamiltonian (\ref{eq:Ham1}) evaluated directly from the HTE.
The structure constants are
(A) $D=11$K, $J_1=10$K and $J=0.5$K; 
(B) $D=19.5$K, $J_1=13.5$K and $J=0.5$K; 
(C) $D=18$K, $J_1=9$K and $J=0.5$K; 
(D) $D=11$K, $J_1=0$K and $J=0.5$K. 
In each case  $\mu_B=0.672$K/T and $g=2.0$.
The critical fields estimated from the magnetization curve coincide with 
the TBA results. 
}
\label{fig:HTEfig1}
\end{figure}

\begin{figure}[tbp]
\begin{center}
\vspace{10mm}
\includegraphics[width=0.70\linewidth]{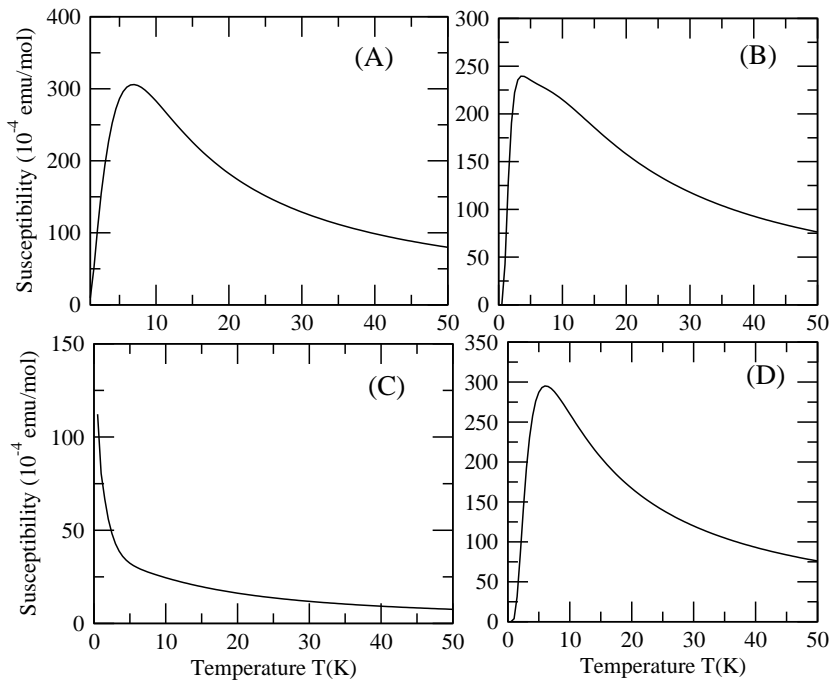}
\end{center}
\caption{Susceptibility versus temperature at $H=0$ T. 
The numerical values are the same as in the previous figure. 
The susceptibility curves in (A), (B) and (D) indicate the existence of 
energy gaps, while no significant gap can be observed in (C).}
\label{FIG2}
\end{figure}

We plot the magnetization versus magnetic field and the susceptibility versus
temperature for different configurations in \Fref{fig:HTEfig1} and \Fref{FIG2}, 
corresponding to the parameter sets found in the TBA analysis in \sref{sec:tba}. 
Comparing the results of both approaches (see \Fref{fig:HTEfig1} for numerical TBA values) 
we find that the critical fields obtained by the TBA coincide well with the values
from the HTE free energy (\ref{eq:freeenergyHTE}), even at relatively low
temperatures.  
For case {(A)} in \Fref{fig:HTEfig1} and
\Fref{FIG2}, where $J_1<D<J_1+4J$, a field-induced gap exists in the
lowest magnon excitation. 
The ground state lies in a magnetic doublet state with an energy gap. 
Note the typical rounded peak in the susceptibility for the antiferromagnetic gapped spin chain.
For case {(B)}, further enhancement of the
anisotropy $D$ into the region $J_1+4J<D<2J_1-4J$ opens a non-magnetic
singlet mid-plateau. 
This leads to a spin-Peierls-like quantum phase transition. 
The magnetic ground state is gapped, as before.  
If the single-ion anisotropy $D$ is large enough, the gapped ground state is
weakened by the singlet potential.  
If $2J_1-4J<D<2J_1+4J$, the ground state shifts into a gapless phase, 
as shown in case {(C)}.  
The \label{page:strangeHaldanegap}
round peak in the susceptibility curve, indicating an energy gap, now gets
replaced by  a steep increase for temperatures close to zero,
indicating a gapless phase. 
In this case, the mid-plateau remains open
if the anisotropy is large enough, i.e., if $2J_1-4J<D<2J_1+4J$. 
The most common gapped  phase is found in case {(D)}, where $J_1=0$ and
the anisotropy is large, $D>4J$. 
A non-magnetic singlet occupies the whole ground state as long as the energy gap 
$\Delta=D-4J$ remains closed by the external magnetic field.
We are now ready to move onto the direct comparison with experimental results.

\section{Examination of real compounds} 
\label{sec:Exp_comp}
\label{sec:exp}

The compounds to be examined via the integrable
spin-$1$ model \eref{eq:Ham1} are listed in Table 1. 
We summarize the coupling constants obtained from the Heisenberg chain (HC) and our 
model (HTE and TBA). 
We shall see below  that the thermal and magnetic properties derived from the HTE
method  are in  excellent agreement with the experimental results.

\begin{center}
\begin{table}
\begin{tabular}{|c|c|c|c|c|c|c|c|c|}
\hline 
 Compound&HC $J$&HC $D$&HC $E$&HTE $J$&HTE $D$&HTE $E$&TBA$H_{c1}$&TBA$H_{c2}$ \\
\hline 
 NiSnCl  &$0.02$K&$0.64$K&$0$K&$0.02$K& $0.64$K&$0$K&$0.38$ T&$ 0.49$ T
\\
\hline
NENC&$0.65$K&$6.15$K&$0.7$K&$0.17$K& $6.4$K&$0$K&$3.8 $ T &$ 4.7$ T
\\
\hline
NBYC&$0.2$K&$2.55$K&$1.5$K&$0.35$K& $2.62$K&$1.49$K& N/A &$ 2.7$ T
\\\hline
NDPK&$0.96$K&$5.0$K&$2.8$K&$ 0.55 $K& $ 5.15$K&$2.55$K&$1.0$ T&$ 4.66$T
\\
\hline 
\end{tabular}
\caption{Estimates for the exchange coupling $J$ and anisotropy constants  
$D$ and $E$ for some real compounds.
Here HC denotes the values obtained from numerical fits for the spin-1 
Heisenberg chain \cite{NENC,sus,ESR,NBYC} and HTE denotes the values 
obtained using the integrable spin-1 model and the HTE method.
Also shown are the corresponding TBA estimates for the critical magnetic fields.
 }
\end{table}
\end{center}

\subsection{Nickel salts}
\label{sec:Nisalt}

Certain Nickel salts can be described
by the spin-1 Hamiltonian \eref{eq:Ham1} with chemical potentials
\eref{eq:Chempotparallel} and \eref{eq:Chempotperpendicular} with $E=0$. 
The chemical potential terms are explicitly given in \eref{eq:chempot2} for the parallel 
case \eref{eq:Chempotparallel}, i.e.  
\begin{eqnarray}
\mu_1&=-\mu_B g H, \nonumber\\
\mu_2&= -D,\nonumber\\
\mu_3&= \mu_B g H,\nonumber
\end{eqnarray}
and for the perpendicular magnetic field case \eref{eq:Chempotperpendicular},
\begin{eqnarray}
\mu_1=-\frac{1}{2}(D+\sqrt{D^2 + (2 \mu_B g H)^2}),\nonumber\\
\mu_2=0,\nonumber\\
\mu_3= -\frac{1}{2}(D-\sqrt{D^2 + (2 \mu_B g H)^2}). \nonumber
\end{eqnarray}

A large zero-field splitting effect was identified in some nickel
salts, such as NiSnCl$_6\cdot 6$H$_2$O (abbreviated as NiSnCl) \cite{PRB3488},
[Ni(C$_5$H$_5$NO)$_6$](ClO$_4$)$_2$ \cite{PRB3523} and 
Ni(NO$_3$)$_2\cdot 6$H$_2$O \cite{PRB4009}. 
Here the Ni$^{2+}$ ion causes a large splitting
between the singlet and doublet in the spin triplet in
an axially distorted crystalline field. 
For $D>0$ and no magnetic field, the triplet is split into a lower
singlet and an excited doublet. 
An early theoretical prediction
for this kind of compound was done as a first order
approximation via Van Vleck's equation \cite{Carlin}. 
In this treatment the exchange interaction is neglected in first order for the specific heat,
magnetization and susceptibility. 
An effective field was introduced  for the magnetization and susceptibility to
fit the experimental data.
In section \ref{sec:tba}, 
we have shown  that large anisotropies $D$ may drive the
antiferromagnetic spin chain into a  gapped phase, which is
significantly different from the valence-bond-solid ground state. 
Examining the magnetic properties of the salt NiSnCl$_6\cdot 6$H$_2$O
\cite{PRB3488} via the HTE method, we find that the measurements are best fitted by a model 
with zero-field splitting $D\approx0.64K$ and a much weaker spin  exchange
interaction $J\approx 0.02K$.
Actually even the zero order HTE with free energy \eref{eq:freeenergyHTE} gives 
a good analytic result for this compound \cite{PRB3488}. 
As expected, the zero order term of the HTE free energy is just the 
result for $L$ non-interacting identical spins with three energy levels 
$\mu_1$, $\mu_2$ and $\mu_3$, i.e.
 \begin{equation}
f(H,T)_{\textrm{\tiny free 
spin}}=-T\ln(e^{-{\mu_1}/{T}}+e^{-{\mu_2}/{T}}+e^{-{\mu_3}/{T}}).
\label{eq:independentsystem}
\end{equation}
Comparing this with the zero order HTE term in \eref{eq:freeenergyHTE}, i.e., with
$\ln Q^{(1)}_1$, we see that \eref{eq:independentsystem}
is identical to the first term in the Q-System \eref{eq:freeenergyHTE}.

\begin{figure}[t]
\begin{center}
\vspace{5mm}
\includegraphics[width=0.6\linewidth]{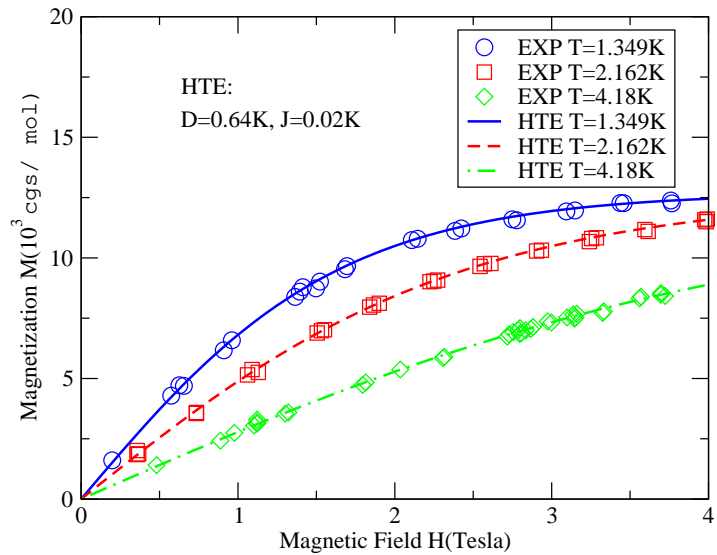}
\end{center}
\caption{Comparison between theory and experiment \cite{PRB3488} for the 
magnetization versus magnetic field of the compound  NiSnCl$_6$$\cdot 6$H$_2$O.
The magnetic field is parallel to the trigonal axis.
A parameter fit suggests the coupling constants $D=0.64$ K, $J=0.02$ K, 
$g_{\parallel}=2.22$ with $\mu_B=0.672$ K/T. 
The conversion constant is
$M_{{\rm HTE}}\approx 8.5 M_{{\rm EXP}}$ ($10^3$ cgs/mol). 
The agreement between the theoretical curves derived here from the HTE approach 
and the experimental data suggests that the compound 
might exhibit zero-field splitting with much stronger  splitting strength $D$ than the
spin exchange interaction $J$. 
}
\label{3488Mag}
\vspace{0.8cm}
\end{figure} 

In \Fref{3488Mag} we compare the magnetization obtained from the 
theoretical free energy expression \eref{eq:freeenergyHTE} with parameters
\begin{eqnarray}
 D \approx 0.64K, \qquad
 J \approx 0.02K, \qquad
 g_{\parallel}=2.22,\nonumber
\end{eqnarray}
at temperatures $T=1.349K$, $T=2.162K$ and $T=4.18K$ with the experimental 
data for NiSnCl$_6$$\cdot 6$H$_2$O.
They show excellent agreement.
The experimental curves are for the case of the external magnetic field parallel 
to the trigonal axis given by the crystal sample. 
The magnetic properties are anisotropic with respect to the external field direction.  
From our TBA analysis, at zero temperature, the energy gap is $\Delta \approx 0.55$ K for this salt. 
At very low temperatures the system mainly occupies the non-degenerate 
singlet ground state if the magnetic field is less than the critical field $H_{c1}$.
The length of the antiferromagnetic correlations is finite.  
For higher temperatures $T>D$, the singlet ground state no longer
dominates, with excitations caused by spin flips.

\begin{figure}[t]
\begin{center} 
\vspace{5mm}
\includegraphics[width=.6\linewidth]{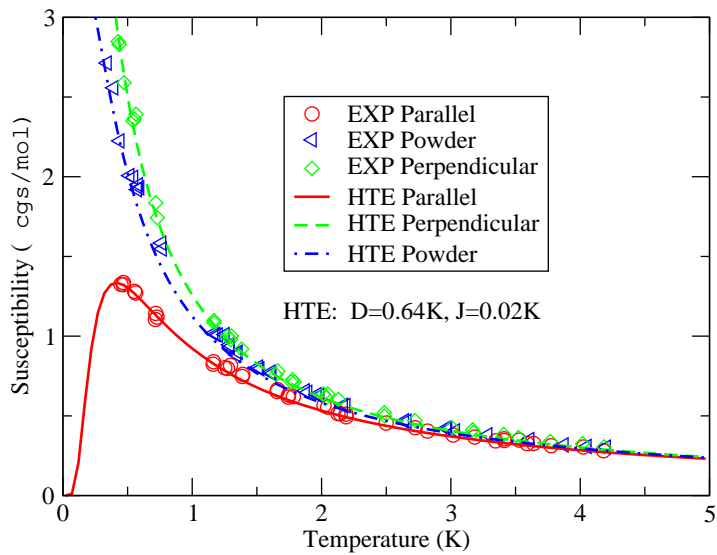}
\end{center}
\caption{Comparison between theory and experiment \cite{PRB3488} for the 
susceptibility  versus temperature of the compound
NiSnCl$_6$$\cdot$$ 6$H$_2$O.  
A parameter fit suggests the coupling constants 
$D=0.64$ K, $J=0.02$ K, $g_{\parallel}= 2.22$ and  $g_{\perp}= 2.20$ with 
$\mu_B=0.672$ K/T. 
The conversion constant is $\chi_{{\rm HTE}}\approx 0.8123 \chi_{{\rm EXP}}$ (cgs/mol). 
The  agreement between the HTE curve and the experimental data confirm that 
the compound NiSnCl$_6$$\cdot $$6$H$_2$O exhibits zero-field splitting with a
larger  strength than the spin exchange interaction.}
\label{fig:Nisaltsusc}
\end{figure} 
 
In some experiments the samples are not of macroscopic crystal size,
thus it is not possible to align the material parallel to an external field
direction. 
In such `powder' samples we assume an isotropic distribution and use the heuristic 
formula (see p$121$ in Ref.~\cite{Carlin}) 
\be 
\chi_{{\rm powder}}\approx \case{1}{3}\chi_{\parallel}+\case{2}{3}\chi_{\perp} ,
\ee
where $\chi_{\parallel}$ and $\chi_{\perp}$ refer to the model Hamiltonian
\eref{eq:Ham1} with respective on-site interactions \eref{eq:Chempotparallel} and
\eref{eq:Chempotperpendicular}.  
This simple ansatz describes the experimental data quite well, as can be seen in
\Fref{fig:Nisaltsusc} where all three susceptibilities -- parallel,
perpendicular and powder -- have been measured for this compound.  
The temperature dependence of the magnetic susceptibility for powdered
samples and the HTE predictions using the above averaging formula are
given in \Fref{fig:Nisaltsusc}. 
The typical rounded peak in the low magnetic field susceptibility curve, 
characteristic of low-dimensional antiferromagnets, are predicted to occur 
around $0.41$ K from the HTE relation 
$\chi_{\parallel}=-\frac{\partial^2}{\partial H^2}f_{HTE}(T,H)$ 
for the parallel magnetic field.  
Using the same physical model, i.e., with the same parameters as for the specific heat
comparison, we find that the susceptibility $\chi _{\parallel}$ is in
good agreement with the experimental curve. 
If the magnetic field is perpendicular to the trigonal axis 
(respectively along the $x$ or $y$ axis ), the susceptibility is given by 
$\chi_{\perp}=-\frac{\partial^2}{\partial H^2}f_{HTE}(T,H)$ 
with the perpendicular magnetic field chemical potentials.  
We see that our theoretical curve, depicted by the dashed line in
\Fref{fig:Nisaltsusc}, gives a good fit to the experimental
susceptibility curve with perpendicular field. 
It also suggests the same parameter constants, except $g_{\perp}=2.20$.

\begin{figure}[t]
\vspace{.3cm}
\begin{center}
\vspace{5mm}
\includegraphics[width=0.6\linewidth]{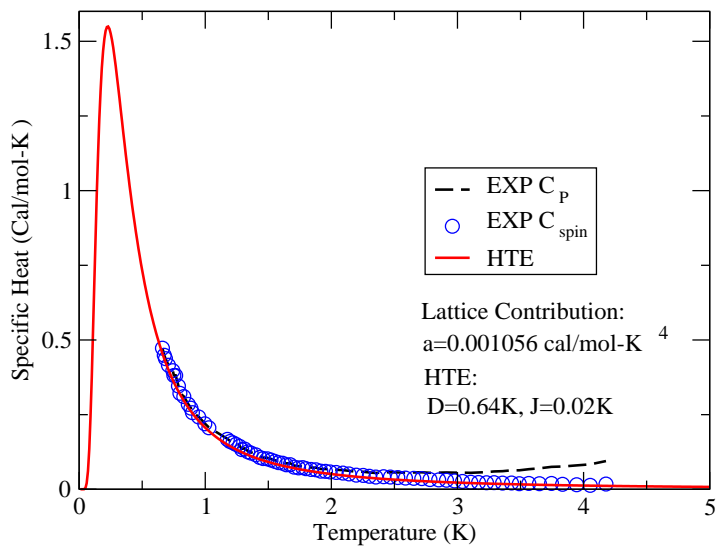}
\end{center}
\caption{Comparison between theory and experiment \cite{PRB3488} for the 
magnetic specific heat versus temperature
of the  compound NiSnCl$_6$$\cdot $$ 6$H$_2$O. 
A parameter fit suggests the coupling constants $D=0.64$ K, $J=0.02$ K, $g= 2.2$ with 
$\mu_B=0.672$ K/T. 
The conversion constant is 
$C_{{\rm HTE}}\approx 1.920 C_{{\rm EXP}} {\rm (cal/mol-K)} \approx 8 C_{{\rm EXP}}$ (J/mol-K).  
The dashed line represents the experimental measured values 
$C_P$ including the lattice contribution. 
The circle curve is the corrected magnetic specific heat without a
lattice contribution of $aT^3$ from $C_P$, where $a \approx
1.056\times 10^{-3 }$ cal/mol-K$^4$.}
\label{fig:Nisaltspecheat}
\end{figure}

In \Fref{fig:Nisaltspecheat}  the specific heat curve for $H=0$ T shows
that the HTE result agrees with the experimental data
\cite{PRB3488} in the available temperature region. 
Note that we do not change the coupling constants to get a best 
fit for each physical property. 
Rather we assign fixed coupling constants 
to each compound and thus truly predict the physical behaviour. 
The dashed line denotes the raw experimental magnetic specific heat
$C_P$ while the circles denote the experimental data, already
corrected for a lattice contribution $aT^3$ with $a \approx 1.056
\times 10^{-3}$ cal/mol K$^4$ \cite{PRB3488}.
In the absence of a magnetic field a rounded peak, indicating short
range ordering, is predicted around $0.23$ K from the HTE result.
Below $T\approx0.23$ K, there is an exponential decay due to an ordered phase.

\subsection{The compound NENC}
\label{sec:NENC}

\begin{figure}[t]
\begin{center}
\vspace{14mm}
\includegraphics[width=0.6\linewidth]{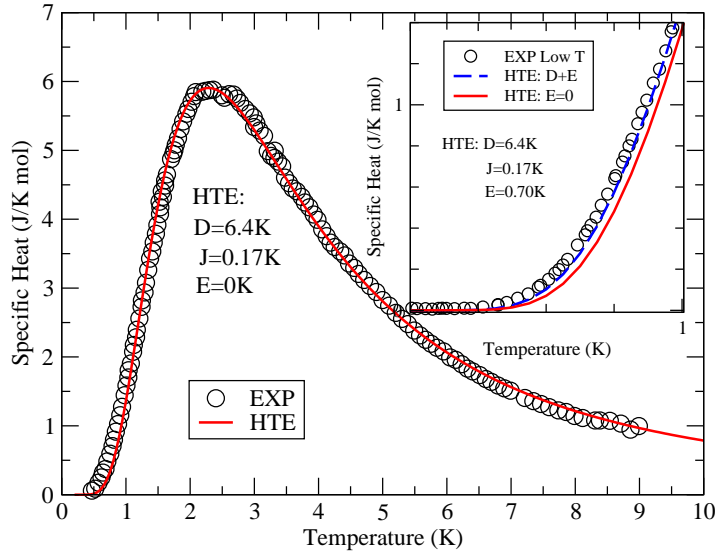}
\end{center}
\caption{Comparison between theory and experiment  \cite{NENC} for the 
magnetic specific heat versus temperature
of the compound
Ni(C$_2$H$_8$)$_2$Ni(CN)$_4$ (NENC). 
The conversion constant  is $C_{{\rm HTE}} \approx 8 C_{{\rm EXP}}$ (J/mol-K).   
The solid line denotes the specific heat  evaluated directly from the free energy  
with the  
parameters $J=0.17$ K, $D=6.4$ K, $E=0$ with $g=2.24$
and $\mu_B=0.672$ K/T. 
The low temperature specific heat curves are shown in the inset.
For low temperatures the
inclusion of in-plane rhombic anisotropy $E=0.7$ K (dashed line) gives a better 
fit than without rhombic anisotropy (solid line).}
\label{FIGheat}
\end{figure}

It is known that antiferromagnetic spin-$1$ chains \cite{NINO1,SP1C1,SP1C2} 
with weak planar anisotropy can exhibit a non-magnetic Haldane phase.  
The large $D$ gapped phase has been observed in the compounds 
NENC, NDPK and NBYC \cite{NENC,NBYC}.
In these compounds the in-plane anisotropy $x^2-y^2$ breaks the $z^2$
symmetry and weakens the planar anisotropy effect.
It was inferred from the experimental analysis in NENC \cite{NENC} 
that the contribution $E$ from the in-plane anisotropy 
is negligible in comparison with large $D$,
where the Nickel(II) $z^2$ orbit along the c-axis forms a strong
crystalline field.
As a result this strong crystalline field dominates
the low temperature physics.
The antiferromagnetic intrachain exchange interaction further lowers 
the energy but its contribution to the ground state, as well as the 
low-lying excitations, is minimal. 
As a consequence, the model (\ref{eq:Ham1}) with
appropriate single-ion anisotropy is expected to describe this
compound rather well.  
The specific heat has been measured up to a temperature around $10$ K \cite{NENC}. 
A typical round peak for short range ordering at $T\approx 2.4$ K is observed
(see \Fref{FIGheat}). 
An exponential decay is detected for temperatures below approx. $2.4$ K. 
The calculated HTE specific heat evaluated from the model (\ref{eq:Ham1}) 
with $J=0.17$ K, $D=6.4$ K (the solid line with $E=0$ in \Fref{FIGheat}) 
is in excellent agreement with the experimental curve in the temperature 
region $T>0.8$ K. 
Our analytic result for the specific heat gives a better fit with the 
experimental data than the perturbation theory prediction \cite{NENC}.  
For low temperatures (below $0.8$ K) paramagnetic impurities and a small
rhombic distortion are the main reasons for the discrepancy.  
The inset of \Fref{FIGheat} shows that the addition of a small rhombic anisotropy
$E=0.7$ K (dashed line) gives a better fit at low temperatures than
that with $E=0$ (solid line).
However, it is negligible for high temperatures.

\begin{figure}[t]
\begin{center}
\vspace{5mm}
\includegraphics[width=.60\linewidth]{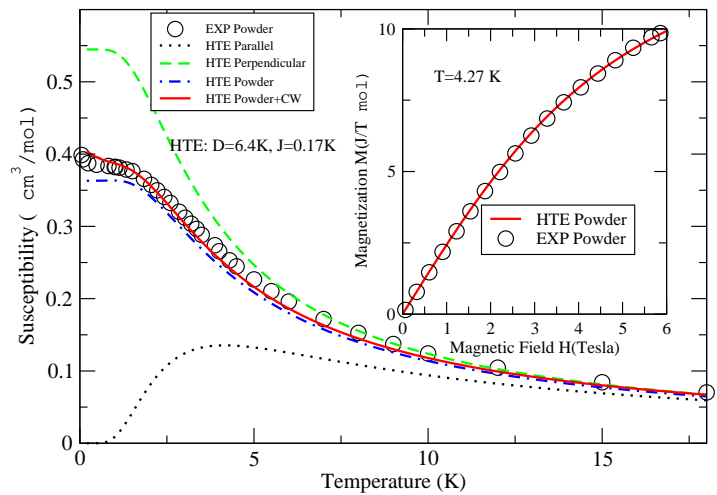}
\end{center}
\caption{
Comparison between theory and experiment \cite{sus} for the 
susceptibility versus temperature of the compound NENC.
The theoretical curve (solid line) is obtained via the empirical relation 
$\chi _{{\rm Powder}}=\frac{1}{3}\chi_{\parallel}+\frac{2}{3}\chi_{\perp}$ 
and by adding a Curie-Weiss contribution. 
The inset shows the comparison for magnetization versus magnetic field at $T=4.27$ K.
The theoretical curve follows from the empirical expression for the powder magnetization.  
A fit for the susceptibility and magnetization suggests the coupling constants 
$J=0.17$ K, $D=6.4$ K and $E=0$, with $g_{\perp}=2.18$ and $g_{\parallel}=2.24 $. 
The conversion constants are the same as in the previous figure \Fref{fig:Nisaltsusc}.}
\label{fig:NENCsusc}
\end{figure}

As far as we know the susceptibility measurement for this compound
was made on powdered samples in a small magnetic field of $H=0.1 $ mT.
The electron spin resonance for this compound was studied in Ref.~\cite{ESR}. 
In \Fref{fig:NENCsusc}, we compare the experimental results with our theoretical
predictions. 
The susceptibility of NENC was measured only in the temperature range 50 mK - 18 K.  
From the data provided in Ref.~\cite{NENC} we could not accurately estimate constants 
for the Curie-Weiss term or the paramagnetic impurities.
The theoretical HTE curves for the susceptibility with  both parallel and 
perpendicular magnetic field are again evaluated from the 
free energy \eref{eq:freeenergyHTE} with chemical potential terms \eref{eq:Chempotparallel} and \eref{eq:Chempotperpendicular}.
We obtained the powder susceptibility  from the heuristic average
\be
\chi _{{\rm Powder}}
=\case{1}{3}\chi_{\parallel}+\case{2}{3}\chi_{\perp}.
\ee
In this form it does not fit the experimental data very well at low temperatures, 
probably due to small contributions from a Curie-Weiss term and
paramagnetic impurities.
We have corrected the powder susceptibility $\chi _{{\rm Powder}}$ with the 
Curie-Weiss contribution $c/(T-\theta)$ to obtain a better agreement with the
experimental curves. 
Here $c \approx 0.045 $ cm$^{3}$ K/mol and $\theta \approx -0.9$ K. 
For our spin chain model, the 
CW-corrected experimental data suggests the 
fitted coupling constants $J=0.17$ K, $D=6.4$ K and $E=0$, with 
$g_{\perp}= 2.18$ and $g_{\parallel}=2.24$. 
In this case the TBA analysis predicts  an energy gap $\Delta \approx 5.72 $ K for a parallel 
external magnetic field.
The inset of \Fref{fig:NENCsusc} shows the magnetization of a powder sample at $T=4.27$ K. 
The singlet is supressed by the temperature.
The best agreement between the experimental data and our powdered magnetization, 
$M_{\rm Powder}=\frac{1}{3}M_{\parallel}+\frac{2}{3}M_{\perp}$,
is found with the anisotropic Lande factors  $g_{\perp}= 2.18$ and $g_{\parallel}=2.24 $. 
Here $M_{\parallel}$ and $M_{\perp}$ denote the magnetization for the external field 
parallel and perpendicular to the axis of quantization.

\subsection{The compound NBYC}
\label{sec:NBYC}

In this section we discuss the application of our theoretical results to
measurements done on the spin-$1$ compound \cite{NBYC}
Ni(C$_{10}$H$_8$N$_2$)$_2$Ni(CN)$_4$ $\cdot$ H$_2$O (NBYC).  
For this compound it is generally believed that the in-plane anisotropy $E$ and the
large anisotropy $D$ play a significant role and so we expect the integrable 
model (\ref{eq:Ham1})
with suitable on-site anisotropy interactions to be a good
microscopic model for this material.
 
The thermodynamic properties have recently been experimentally investigated
in Ref.~\cite{NBYC}. 
Theoretical studies  based on a strong-coupling expansion method \cite{Spathis} 
suggest that the anisotropy of this compound might lie close to the boundary between
the observed Haldane and field-induced gapped phases \cite{NBYC}. 
However, due to the validity of the strong-coupling expansion method \cite{NBYC}, 
there are apparent descrepancies in fitting the specific heat, susceptibility and 
magnetization if the rhombic anisotropy is large.
Here we evaluate the relevant quantities for the powdered samples
using the free energy expression
\eref{eq:freeenergyHTE} with parallel and perpendicular magnetic fields,
in-plane anisotropy $E$ and single-ion anisotropy $D$ 
(see \eref{eq:Chempotparallel} and \eref{eq:Chempotperpendicular} 
and their eigenvalues \eref{eq:EV.1}-(\ref{eq:EV.6})) 
via the empirical formula 
$\chi_{{\rm Powder}} =\frac{1}{3}\chi_{\parallel}+\frac{2}{3}\chi_{\perp}$.

\begin{figure}[t]
\begin{center}
\vspace{5mm}
\includegraphics[width=0.6\linewidth]{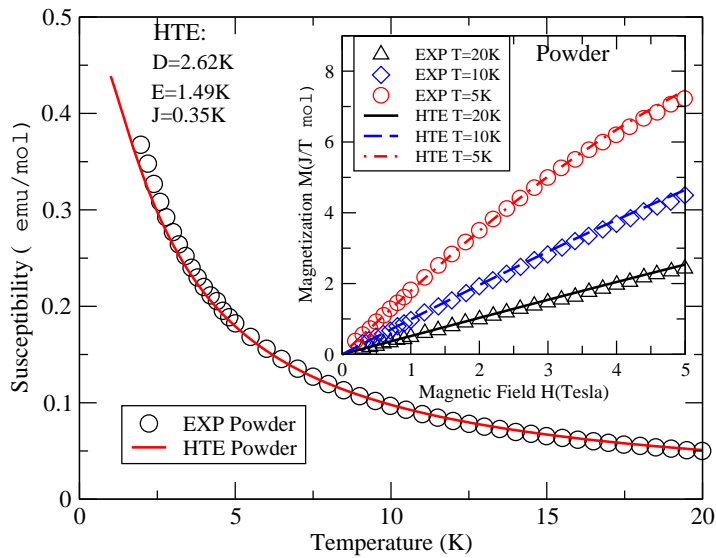}
\end{center}
\caption{Comparison between theory and experiment \cite{NBYC} for the
susceptibility versus temperature of the spin-$1$ chain compound  NBYC. 
The conversion constants are the same as for the above materials.
The solid line is the susceptibility evaluated directly
from the free energy of the spin-$1$ chain for the powdered samples with
coupling constants $J=0.35 $ K, $D=2.62$ K and $E=1.49$ K,
with $g_{\parallel}=g_{\perp}=2.05$.
The small discrepancy at low temperature might be attributed to a Cuire-Wess
contribution.
The inset  shows the magnetization for the powder samples at T=$5,10,20$ K. 
}
\label{fig:NBYC}
\end{figure}

In \Fref{fig:NBYC} we compare the results for the susceptibility of NBYC.
A fit for the susceptibility suggests the coupling strengths 
$D=2.62$ K, $E=1.49$ K and $J=0.35 $ K, with $g_{\parallel}=g_{\perp}=2.05$.  
The small discrepancy at low temperature is probably caused by a Curie-Weiss 
contribution. 
The inset of \Fref{fig:NBYC} shows plots of the magnetization of the 
powdered samples at T=$5,10,20$ K.  
Again our theoretical curves are evaluated from the empirical relation 
$M_{\rm Powder}=\frac{1}{3}M_{\parallel}+\frac{2}{3}M_{\perp}$. 
An overall agreement for the magnetization at different temperatures is consistent 
with the parameters used for the susceptibility. 
The singlet state is now supressed by both the in-plane rhombic anisotropy and the 
temperature.

\begin{figure}[t]
\begin{center}
\vspace{14mm}
\includegraphics[width=.60\linewidth]{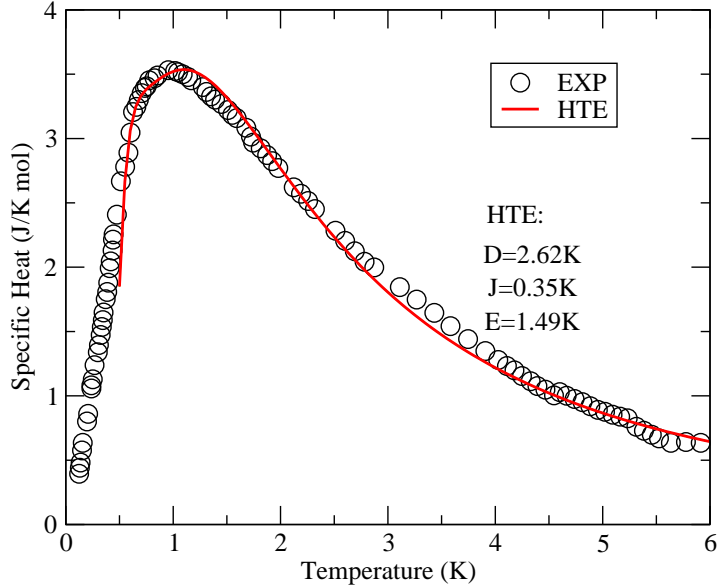}
\end{center}
\caption{Comparison between theory and experiment \cite{NBYC} for the magnetic specific heat 
versus temperature of the compound NBYC. 
The conversion constant is $C_{{\rm HTE}} \approx 10 C_{{\rm EXP}}$ (J/mol-K). 
The solid line denotes the specific heat at $H=0.1$ mT evaluated directly 
from the free energy with parameters $D=2.62$ K, $E=1.49$ K and
$J=0.35 $ K,   with $g=2.05$ and $\mu_B=0.672$ K/T.}
\label{fig:NBYCheat}
\end{figure}

We turn now to the specific heat which has been measured up to a temperature 
of $6$ K \cite{NBYC}. 
The theoretical specific heat evaluated from the integrable model (\ref{eq:Ham1}),
 with the same parameters given above, 
 is in good  agreement with
the experimental data in the temperature region $0.5$ K to $6$ K, 
as can be seen in \Fref{fig:NBYCheat}.
For lower temperatures the HTE does not give reliable results due to 
poor convergence.

\subsection{The compound NDPK}
\label{sec:NDPK}

\begin{figure}
\begin{center}
\vspace{10mm}
\includegraphics[width=.60\linewidth]{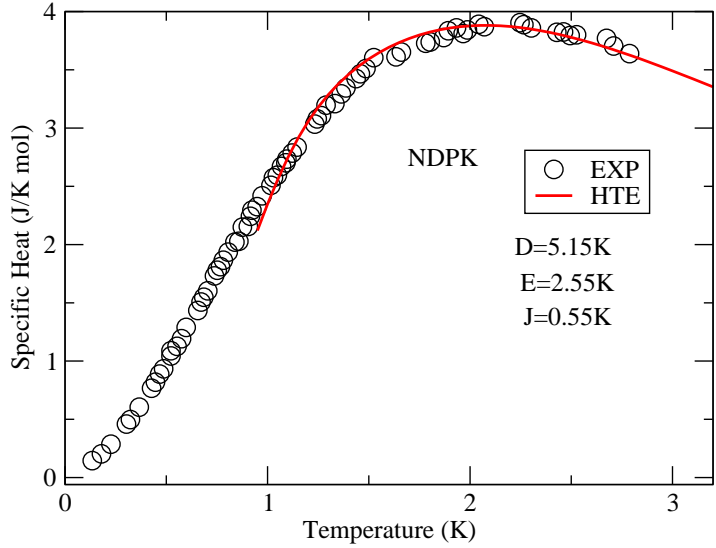}
\end{center}
\caption{Comparison between theory and experiment \cite{NENC} for the
magnetic specific heat versus temperature of the compound  NDPK. 
The conversion constant is $C_{{\rm HTE}} \approx 10 C_{{\rm EXP}}$ (J/mol-K). 
The solid line denotes the specific heat at $H=0$ T evaluated directly from 
the free energy  with the  parameters  $D=5.15$ K, $E=2.55$ K and
$J=0.55 $ K.}
\label{fig:NDPKheat}
\end{figure}

Another compound considered in Ref.~\cite{NENC} to be a spin-$1$
magnetic chain with planar and in-plane anisotropy is NDPK.
As far as we are aware, there has been no comprehensive theoretical and
experimental study of this compound. 
The specific heat of  NDPK was measured only in the temperature range 
from $100$ mK to $2.5$ K. 
A numerical {ESCA} study \cite{NENC} of the specific heat indicated
this compound lies in the large-$D$ phase. 
Quantitative agreement with the experimental data suggested the values 
$D=5$ K, $J=0.96$ K and $E=2.8$ K \cite{NENC}. 
We examined the specific heat via the integrable model (\ref{eq:Ham1}) with the 
parameters $D=5.15$ K, $J=0.55$ K and $E=2.55$ K.
The specific heat (solid line ) evaluated
from the HTE is in good agreement with experimental data in the
temperature region $0.9$ K to $2.5$ K (see \Fref{fig:NDPKheat}). 
For  temperatures below $1$ K, the HTE is not valid due to the small ratio of $J/T$.  
We believe that the integrable model will  also describe  other magnetic properties 
of NDPK.

\section{The integrable spin-$\frac32$ chain}
\label{sec:spin32}

\subsection{The exactly solved $su(4)$ model}

It is generally believed that the ground state of the antiferromagnetic spin-$S$ 
Heisenberg chain 
remains in a gapless phase when $2S$ is an odd integer \cite{Hald,AFF1}.  
Nevertheless, an appropriate projection operator can open a 
``Haldane gapped phase'' \cite{AFF3}.  
Numerical study \cite{SP4} of the spin-$\frac32$ antiferromagnetic chain has verified the 
existence of a one third saturation magnetization plateau. 
In this section we consider an integrable spin-$\frac32$ chain with two different kinds of 
single-ion anisotropies, capable of triggering gapped and gapless phases.
We investigate the thermodynamics of the model in terms of the HTE and TBA.  
The two Hamiltonians in the fundamental basis are given by 
\begin{equation}
{\cal H}=J\sum_{j=1}^{L}P_{j,j+1}^{(3)}+D\sum_{j=1}^LP_j^{\pm}-\mu_Bg H\sum_{j=1}^LS^z_j, \label{Ham2}
\end{equation}
where 
\begin{equation}
P^{(3)}_{j,j+1}=\sum_{\alpha=0}^{2}(-1)^{2-\alpha}\prod_{\beta \neq
 \alpha}^{2}\frac{\vec{S}_j\cdot\vec{S}_{j+1}-x_{\beta}}{x_{\alpha}-x_{\beta}},
\end{equation}
with $x_{\alpha}=\frac{1}{2}\alpha (\alpha+1)-S(S+1)$ \cite{classify}.  
We define model 1 with projection operator $P^+$ and model 2 with $P^-$, where
\begin{eqnarray}
P_j^{+}&=&(S^z_j+\case{3}{2})(S^z_j-\case{3}{2})(S^z_j+\case{1}{2}),
\nonumber\\
P_j^{-}&=&(S^z_j)^2.
\end{eqnarray}
In the above, $\vec{S}_j$ is the spin-$\frac32$ operator acting on  site $j$, 
$J$ is the exchange coupling and $D$ is the single-ion anisotropy. 
The operator $P^+_j$ projects spin states onto the state with  $M^z=\frac12$, which leads to a
magnetic ground state.  
In general, one can incorporate different projection operators in the spin-$S$ chain 
associated with the integrable $su(n)$ model with $n=2S+1$ \cite{sun} such that 
multi-plateaux phases can occur.

\subsection{HTE approach}

To investigate the thermodynamic properties we again apply the HTE scheme. 
The eigenvalue of the QTM (up to a constant) is \cite{Fujii,ZT}
\begin{eqnarray}
T^{(1)}_1(v,\left\{v^{(a)}_i\right\})&=&\e^{-\beta \mu_1}\phi _-(v-\mathrm{i})
\phi _+(v)\frac{Q_1(v+\frac12\mathrm{i})}{Q_1(v-\frac12\mathrm{i})}\nonumber\\
& &
+\e^{-\beta \mu_2}\phi _-(v)\phi _+(v)
\frac{Q_1(v-\frac32\mathrm{i})Q_2(v)}{Q_1(v-\frac12\mathrm{i})Q_2(v-\mathrm{i})}\nonumber\\
& &+\e^{-\beta \mu_3}\phi _-(v)\phi _+(v)
\frac{Q_2(v-2\mathrm{i})Q_3(v-\frac12\mathrm{i})}{Q_2(v-\mathrm{i})Q_3(v-\frac32\mathrm{i})}\nonumber\\
& &
+\e^{-\beta \mu_4}\phi _-(v)\phi _+(v+\mathrm{i})
\frac{Q_3(v-\frac52\mathrm{i})}{Q_3(v-\frac32\mathrm{i})}.\label{EQTM3-2}
\end{eqnarray} 
In the above equations, $\mu_i$, $i=1,2,3,4$, are the eigenvalues of the on-site chemical potentials 
with regard to the fundamental basis $\st[S^z]$
\begin{equation}
\st[1] =\st[\case{3}{2}],\,\st[2] =\st[\case{1}{2}],\,\st[3] =\st[\!-\!\case{1}{2}],\,\st[4] =\st[\!-\!\case{3}{2}].
\label{eq:su4basis}
\end{equation}
For model 1 with projection operator $P^{+}$  the chemical potentials are
\begin{eqnarray}
\mu_1 =& -\case{3}{2}\mu_B gH,\,\,\,\,\,   &  \mu_2 = -\case{1}{2}\mu_B gH-2D,\nonumber\\
\mu_3 =&\case{1}{2}\mu_B gH, \,\,\,\,\,  &  \mu_4=\case{3}{2}\mu_B gH.
\label{eq:chemP1}
\end{eqnarray}
For model 2 with projection operator $P^{-}$ the chemical potentials are
\begin{eqnarray}
\mu_1 =& -\case{3}{2}\mu_B gH,\,\,\,\, &\mu_2 = -\case{1}{2}\mu_B gH-2D,\nonumber\\
\mu_3 =&\case{1}{2}\mu_B gH-2D, \,\,\,\, &\mu_4=\case{3}{2}\mu_B gH.
\label{eq:chemP2}
\end{eqnarray}

The HTE expansion of the free energy, up to third order, is given by
\begin{equation}
-\frac{1}{T}f(T,H)=\ln{Q_{1}^{(1)}}+C^1_{1,0}\left(\frac{J}{T}\right)+C^1_{2,0}\left(
\frac{J}{T}\right)^2+C^1_{3,0}\left(\frac{J}{T}\right)^3+\cdots.\label{eq:EQTM3}
\end{equation} 
where now \cite{ZT}
\begin{eqnarray}
c_{1,0}^{(1)}&=2 \frac{Q^{(2)}_1}{{Q^{(1)}_1}^2}, \nonumber\\
c_{2,0}^{(1)}&=3 \frac{Q^{(2)}_1}{{Q^{(1)}_1}^2}- 6 \frac{{Q^{(2)}_1}^2}{{Q^{(1)}_1}^4} 
+3 \frac{{Q^{(3)}_1}}{{Q^{(1)}_1}^3},\nonumber\\
c_{3,0}^{(1)}&= \frac{10}{3}\frac{Q^{(2)}_1}{{Q^{(1)}_1}^2}
-18\frac{{Q^{(2)}_1}^2}{{Q^{(1)}_1}^4}	 
+\frac{80}{3}\frac{{Q^{(2)}_1}^3}{{Q^{(1)}_1}^6}
+8\frac{{Q^{(3)}_1}}{{Q^{(1)}_1}^3}
-24\frac{{Q^{(2)}_1}{Q^{(3)}_1}}{{Q^{(1)}_1}^5}+ 4 \frac{Q^{(4)}_1}{{Q^{(1)}_1}^4}.
\nonumber 
\end{eqnarray}
Here the Q-system for $su(4)$ is 
\begin{eqnarray}
\fl
Q_1^{(1)} &=& \e^{-\beta \mu_1}+\e^{-\beta \mu_2}+\e^{-\beta \mu_3}+\e^{-\beta \mu_4},\nonumber \\
\fl
Q_1^{(2)} &=& \e^{-\beta\mu_1-\beta\mu_2}+\e^{-\beta\mu_1-\beta \mu_3}+\e^{-\beta\mu_1-\beta \mu_4}+\e^{-\beta \mu_2-\beta\mu_3}+\e^{-\beta \mu_2-\beta\mu_4}+\e^{-\beta \mu_3-\beta\mu_4},
\nonumber \\
\fl
Q_1^{(3)}& =& \e^{-\beta\mu_1 -\beta \mu_2 -\beta \mu_3}+\e^{-\beta\mu_1 -\beta \mu_2 -\beta \mu_4}+\e^{-\beta\mu_1 -\beta \mu_3 -\beta \mu_4}+\e^{-\beta\mu_2 -\beta \mu_3 -\beta \mu_4},\nonumber\\
\fl
Q_1^{(4)} &=& \e^{-\beta\mu_1 -\beta \mu_2 -\beta \mu_3-\beta \mu_4}.\nonumber
\end{eqnarray}

\subsection{TBA analysis: ground state properties and magnetization plateau}

\begin{figure}[t]
\begin{center}
\vspace{5mm}
\includegraphics[width=.750\linewidth]{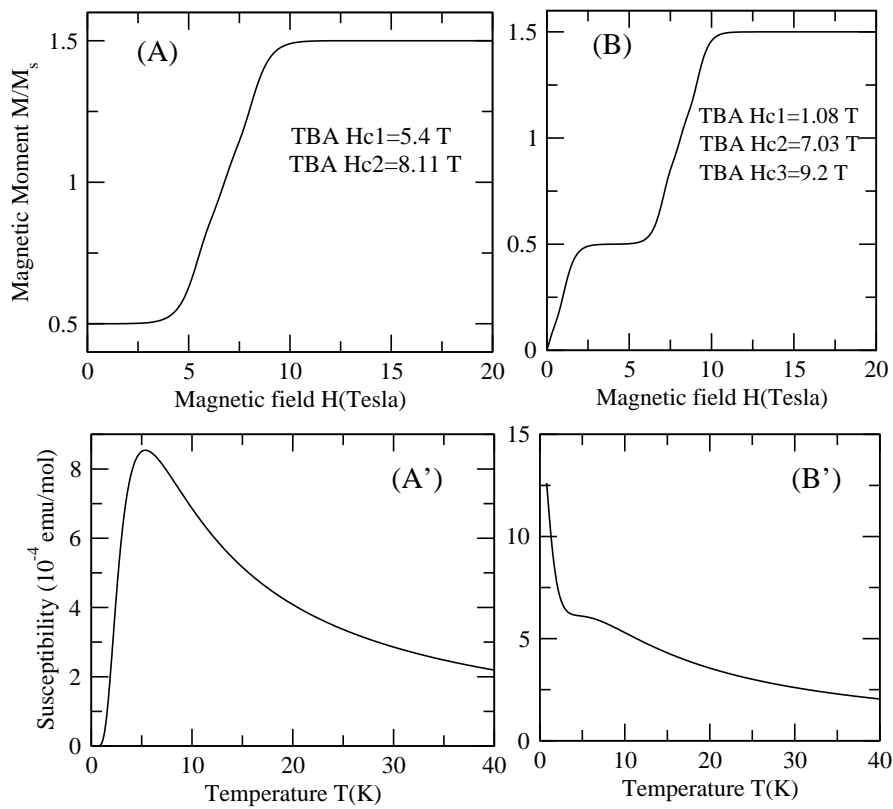}
\caption{
Theoretical curves for the spin-$\frac32$ chain model (\ref{Ham2}) with single-ion anisotropy.
Parts (A) and (B) show the magnetization versus magnetic field at $T=0.8$K. 
In part (A) $D=5$ K and $J=0.5$ K which suggests a magnetic gapped ground state 
for the Hamiltonian (\ref{Ham2}) with projection operator  $P^+$.
In part (B) $D=6$ K and $J=0.4$ K which suggests a gapless ground state for the 
Hamiltonian (\ref{Ham2}) with the single-ion anisotropy $P^-$.
We have set $\mu_B=0.672$ K/T and $g=2.20$. 
The critical fields estimated from the magnetization curves coincide with the TBA results.
Parts (A') and (B') show the susceptibility versus temperature at $H=0$ T. 
The numerical values in {(A')} and {(B')} are the same as in parts {(A)} and {(B)}, respectively. 
The susceptibility curves in (A') indicate the existence of an energy gap, whereas no
significant gap can be observed in figure {(B')}, however, 
the mid-plateau suggests a spin-Peierls-like quantum phase transition.
}
\label{FIG4}
\end{center}
\end{figure}

First consider  Hamiltonian (\ref{Ham2}) with projection operator $P^+$. 
The anisotropy terms $P^{\pm}$ and the external field can be  diagonalized simultaneously
in the fundamental basis \eref{eq:su4basis}.
If $D > 0$, the projection operator $P_j^{+}$ energetically favours 
the component $\st[2]$. 
Proceding in an analogous fashion to the spin-$1$ TBA analysis, 
we find that the magnetic ground state is separated from the lowest magnon 
excitation by the gap $\Delta = 2D-4J$ in the zero temperature limit. 
A  magnetization plateau with $M=\frac12$ starts from zero field. 
The gap is diminished by the magnetic field. 
If the magnetic field is larger than the critical field $H_{c1}=(2D-4J)/\mu_Bg$, 
a phase transition from a ferrimagnetic phase to an antiferromagnetic phase occurs. 
As the magnetic field increases, the magnetization almost linearly increases up to the 
saturation magnetization.
The saturation critical field is $H_{c2}=(2D+4J)/\mu_Bg$.  
The magnetization and the susceptibility evaluated from the free energy (\ref{eq:EQTM3}) 
with the chemical potentials given by \eref{eq:chemP1} are shown in parts
{(A)} and {(A')} of \Fref{FIG4}. 
The magnetization curve indicates that the ground state remains gapped until the magnetic field
exceeds the critical field $H_{c1}$. 
A round peak is found in the susceptibility.  
This conclusion coincides with  the suggestion by Masaki et al. \cite{AFF3} that the  gapped phase  
might exist in the $2S$ odd integer antiferromagnetic spin chains.

Apart from this degenerate ground state, there also exists another
gapped $M=\frac12$ magnetization plateau in the spin-$\frac32$ chain. 
If we consider the Hamiltonian (\ref{Ham2}) with a large anisotropy term
$P^-$, the ground state lies in a gapless phase. 
The gapless ground state could be shifted into the gapped phase with a 
magnetization plateau $M=\frac12$. 
The field-induced magnetic effects occur as follows: 
(i) the system undergoes a phase transition from an antiferromagnetic phase into 
a ferrimagnetic phase at the critical point $H_{c1}=4J/\mu_Bg$; (ii)
the $M=\frac12$ magnetization plateau vanishes at $H_{c2}=(2D-4J)/\mu_Bg$;
(iii) the ground state is fully-polarized beyond the critical field $H_{c3}=(2D+4J)/\mu_Bg$.  
The magnetization curve given in part {(B)} of \Fref{FIG4} indicates the existence
of a mid-plateau, which is reminiscent of the mixed spin-$(1,\frac12)$ ladder \cite{mix}. 
Remarkably, with regard to this mid-plateau, a cusp-like phase transition emerges in the
susceptibility curve in ({B'}) of \Fref{FIG4}. 
This field-induced cusp is similar to the spin-Peierls transition in spin-$\frac12$ chain
materials \cite{SPT}. 
We anticipate that this novel phase transition may exist in real spin-$\frac32$ chain compounds.

\section{Conclusion}
\label{sec:conclusions}

We have studied the thermal and magnetic properties of spin-$1$
and spin-$\frac32$ chains via the $su(3)$ and $su(4)$ integrable models.
This has enabled some novel phase transitions and critical behaviour to be
predicted via the TBA analysis and HTE method.  
In particular, the zero temperature TBA predictions and the temperature-dependent  
analytic result derived from the HTE provide the full ground state properties and 
thermodynamics of the underlying models. 
The integrable spin-$1$ model with appropriate chemical potentials gives
a good description of some real spin-$1$ chain compounds with large single-ion anisotropy. 
The planar anisotropy on-site interaction leads to a gapped phase 
which is significantly different from the Haldane phase in which the spin-spin exchange 
interaction results in a valence bond solid ground state. 
We have compared the theoretical curves with measurements on spin-$1$ chain 
compounds such as the Nickel salt NiSnCl$_6\cdot 6$H$_2$O (\sref{sec:Nisalt}), NENC 
(\sref{sec:NENC}), NBYC (\sref{sec:NBYC}), NDPK (\sref{sec:NDPK}).  
In general, the agreement with the experimental data for the compounds under 
comparison is excellent.
Moreover, a specific anisotropy in the spin-$1$ chain can
trigger a mid-plateau which causes a spin-Peierls-like quantum phase transition. 

We have considered two types of on-site anisotropies  for the integrable spin-$\frac32$ chain.
In contrast to the non-magnetic gapped phase in the spin-$1$ chain, the magnetic
gapped ground state can be separated from the lowest magnon excitation
in the integrable spin-$\frac32$ chain. 
Counting the magneitzation plateau in units of the saturation value $M_s=\frac{3}{2}$, 
a one third magnetization plateau with $M=\frac12$ can start at zero magnetic field.
Moreover, an appropriate single-ion anisotropy can open a one third
mid-magnetization plateau as also exhibited  in the mixed
spin-$(1,\frac12)$ ladder model \cite{mix}.
We anticipate that the one third  magnetization plateau may be observed 
in some real spin-$\frac32$ compounds. 

{\ack 
This work has been supported by the Australian Research Council. 
N. Oelkers also thanks DAAD for financial support. A. Foerster thanks
CNPq and FAPERGS for financial support.
We thank Z. Tsuboi and H.-Q. Zhou for helpful discussions and 
M. Oren\'{a}\u{c} for providing us with experimental
results and helpful correspondence.}

\end{document}